# Closed-form expressions for effective constitutive parameters and electro/magneto-strictive tensors for bi-anisotropic metamaterials and their use in optical force density calculations

Neng Wang[1], Shubo Wang[1,2], Zhao-Qing Zhang[1], and C. T. Chan[1*]

[1]*Department of Physics, The Hong Kong University of Science and Technology, Hong Kong, China.*

[2]*Department of Physics, City University of Hong Kong, Hong Kong, China.*

Correspondence to: phchan@ust.hk

**Abstract**. Using a multiple scattering technique, we derived closed-form expressions for effective constitutive parameters and electro/magneto-strictive tensor components for 2D bi-anisotropic metamaterials. Using the principle of virtual work, we obtained the electromagnetic stress tensor that can be used to calculate the optical force density inside such media. The analytic expressions are tested against full wave numerical simulations. Our effective medium theory is essential for providing a complete macroscopic description of the optical and opto-mechanical properties of bi-anisotropic composites.

Metamaterials are artificial composite materials carrying sub-wavelength inclusions. These materials are designed to have optical or acoustic properties that are not found in nature and hence can be employed to realize novel phenomena that were once thought to be impossible [1-8]. The unusual properties of metamaterials are frequently attributed to their unconventional constitutive parameters, which can be very different from those of the constitutive components. As the properties of metamaterials are characterized by macroscopic constitutive parameters, it is essential to find good effective medium theories that can produce accurate descriptions of the composite material so that we do not need to worry about the complex structural details at the subwavelength level. It is perhaps not an exaggeration to say that effective medium theory is the cornerstone of the concept of metamaterials.

Developing analytic effective medium approximations is often challenging [9-11], particularly for anisotropic composites. Bi-anisotropic metamaterials, which has many special optical properties [12-22], frequently have complex underlying structures and building an effective medium theory for them is not easy. In this paper, we will use analytic techniques to formulate an effective medium approach for 2D bi-anisotropic metamaterials. Our approach is complete in the sense that it not only provides the usual constitutive parameters that can calculate scattering and absorption, but also provides enough information to evaluate microscopic details such as the optical force density inside the metamaterial, which is known to be more difficult to calculate than the total force. Such capability enables us to consider the opto-mechanical response of such materials, in particular soft composites which can be deformed in light fields due to the non-uniform optical force distribution [23-25].

Using multiple scattering theory (MST), we analytically derived the macroscopic effective parameters as well as the electro/magneto-strictive tensors for these materials. Using the virtual work method, we derived the electromagnetic (EM) stress tensor (called extended Helmholtz stress tensor thereafter) for bi-anisotropic media, which can be used to calculate the optical force density inside a bi-anisotropic metamaterial with complex structures. We note that traditional effective medium theories do not provide sufficient information to study the optical force density because they do not give expressions for the electro/magneto-strictive tensor components. By comparing the results produced by different stress tensors, we demonstrate that the extended Helmholtz stress tensor gives the most accurate description of optical force density in bi-anisotropic metamaterials.

The bi-anisotropic metamaterials considered in this study consist of identical chiral inclusions arranged into a regular lattice in the *xy* plane embedded in air with permittivity $\varepsilon_0$ and permeability $\mu_0$, as shown schematically in Fig. 1a. In the long-wavelength limit, the effective relative constitutive parameters of the metamaterials have diagonal matrix forms as

$$\ddot{\varepsilon}_e = \varepsilon_t \hat{x}\hat{x} + \varepsilon_t \hat{y}\hat{y} + \varepsilon_z \hat{z}\hat{z}, \quad \ddot{\mu}_e = \mu_t \hat{x}\hat{x} + \mu_t \hat{y}\hat{y} + \mu_z \hat{z}\hat{z}, \quad \ddot{\kappa}_e = \kappa_t \hat{x}\hat{x} + \kappa_t \hat{y}\hat{y} + \kappa_z \hat{z}\hat{z}. \tag{1}$$

In such a bi-anisotropic material, the EM wave has two eigenmodes with corresponding wave numbers satisfy the following relationships:

$$K_+^2 K_-^2 = k_0^4 (\varepsilon_z \mu_z - \kappa_z^2)(\varepsilon_t \mu_t - \kappa_t^2), \tag{2}$$

where the wave number product can be factorized into two parts involving the out-of-plane (with subscript *z*) and in-plane (with subscript *t*) constitutive parameters, respectively, and $k_0$ is the wave number in air. The dispersion relation for a periodic system can be obtained by considering the secular equation derived from the MST, and in the long-wavelength limit it reduces to (detail derivation is given in supplemental material Sec. I)

$$PK^4 k_0^{-4} + QK^2 k_0^{-2} + RS = 0, \qquad (3)$$

where

$$P = (i+\Lambda A_1)(i+\Lambda B_1) - \Lambda^2 C_1^2, \ R = (1+i\Lambda A_0)(1+i\Lambda B_0) + \Lambda^2 C_0^2, \ S = (i-\Lambda A_1)(i-\Lambda B_1) + \Lambda^2 C_1^2, \quad (4)$$

with $\Lambda = 4/k_0^2 \Omega$ and $\Omega$ is volume of the unit cell. According to the Vieta's formulas, the two solutions to Eq. (3), $K_+^2$ and $K_-^2$, fulfill the relationship

$$K_+^2 K_-^2 = k_0^4 RSP^{-1}. \qquad (5)$$

By comparing Eqs. (2) and (5), we can derive the closed-form expressions of the effective constitutive parameters as (see detail in Supplemental material Sec. I. C [26])

$$\varepsilon_z = 1+i\Lambda B_0, \ \mu_z = 1+i\Lambda A_0, \ \kappa_z = i\Lambda C_0, \quad \varepsilon_t = (i-\Lambda A_1)(i+\Lambda B_1)P^{-1} + \Lambda^2 C_1^2 P^{-1},$$
$$\mu_t = (i+\Lambda A_1)(i-\Lambda B_1)P^{-1} + \Lambda^2 C_1^2 P^{-1}, \qquad \kappa_t = -2i\Lambda C_1 P^{-1}. \qquad (6)$$

For the special case of isotropic inclusions, Eq. (6) reduces to a Maxwell-Garnett form expression whose transverse components bear some resemblance to the 3D chiral Maxwell-Garnett formulas [29, 30], see Supplemental material Sec. I. C [26]. The electro/magneto-strictive tensors can also be obtained using the MST (see Supplemental material Sec. I. D. [26]) after some tedious derivations, which have the following expressions:

$$\frac{\partial \varepsilon_z}{\partial u_{xx}} = 1-\varepsilon_z, \ \frac{\partial \mu_z}{\partial u_{xx}} = 1-\mu_z, \ \frac{\partial \kappa_z}{\partial u_{xx}} = -\kappa_z, \ \frac{\partial \varepsilon_t}{\partial u_{xx}} = -\frac{\varepsilon_t^2 + \kappa_t^2 - 1}{2} + \frac{(\varepsilon_t-1)^2 + \kappa_t^2}{2}\gamma\cos 2\phi_K,$$
$$\frac{\partial \mu_t}{\partial u_{xx}} = -\frac{\mu_t^2 + \kappa_t^2 - 1}{2} + \frac{(\mu_t-1)^2 + \kappa_t^2}{2}\gamma\cos 2\phi_K, \ \frac{\partial \kappa_t}{\partial u_{xx}} = -\frac{(\varepsilon_t+\mu_t)\kappa_t}{2} + \frac{(\varepsilon_t+\mu_t-2)\kappa_t}{2}\gamma\cos 2\phi_K. \quad (7a)$$

$$\frac{\partial \varepsilon_z}{\partial u_{xy}} = \frac{\partial \mu_z}{\partial u_{xy}} = \frac{\partial \kappa_z}{\partial u_{xy}} = 0, \ \frac{\partial \varepsilon_t}{\partial u_{xy}} = -\frac{(\varepsilon_t-1)^2 + \kappa_t^2}{2}\nu\sin 2\phi_K,$$
$$\frac{\partial \mu_t}{\partial u_{xy}} = -\frac{(\mu_t-1)^2 + \kappa_t^2}{2}\nu\sin 2\phi_K, \ \frac{\partial \kappa_t}{\partial u_{xy}} = -\frac{(\varepsilon_t+\mu_t-2)\kappa_t}{2}\nu\sin 2\phi_K. \qquad (7b)$$

where $\gamma = 1.298, \nu = 0.596$ for square, $\gamma = 0.499, \nu = -1.0$ for hexagonal and $\gamma = \nu = 0$ for random lattice structures, $u_{ik} = (\partial u_i/\partial x_k + \partial u_k/\partial x_i)/2$ is the strain tensor with **u** being the displacement vector [31], and $\phi_K$ can be well approximated by the direction of Poynting vector for the effective fields [27]. The tensor components with respect to $u_{yy}$ are obtained by changing the sign of $\gamma$ in Eq. (7a). The electro/magneto-strictive tensors, which depend on the lattice symmetry, describe the stiffness of the constitutive parameters under stretching (diagonal terms) and shearing (off-diagonal terms) [32-35]. They are kernel functions in the extended Helmholtz stress tensor which is useful for calculating the optical force density inside the medium.

The validity of the derived electro/magneto-strictive tensors in Eq. (7) can be tested by comparing with numerical simulation results, which are obtained numerically using eigen-fields and band dispersions combined with finite-difference method (see details in Supplemental material Sec. II. A [26]). The comparison results for both square and hexagonal lattice structures are shown in Fig. 2. The tensor components $\partial \varepsilon_z / \partial u_{xy}, \partial \mu_z / \partial u_{xy}, \partial \kappa_z / \partial u_{xy}$ are zero and not shown here. We can see clearly that the electro/magneto-strictive tensors obtained from numerical calculations (circles) are essentially identical to those calculated using the formulas (lines), indicating that our derived formulas are correct.

According to the virtual work principle [31], the work done by a stress acting on a material boundary is equal to the variation of total EM energy. According to this identity, we can obtain the expression of the extended electromagnetic stress tensor for the bi-anisotropic medium as (see detail in Supplemental material Sec. III [26])

$$T_{ik} = \mathrm{Re}\{\frac{E_i D_k^* + H_i B_k^*}{2} - \frac{1}{4}(\mathbf{E}\cdot\mathbf{D}^* + \mathbf{H}\cdot\mathbf{B}^*)\delta_{ik}$$
$$-\frac{\varepsilon_0}{4}\sum_j \frac{\partial \varepsilon_j}{\partial u_{ik}}|E_j|^2 - \frac{\mu_0}{4}\sum_j \frac{\partial \mu_j}{\partial u_{ik}}|H_j|^2 + \frac{1}{2c}\sum_j \frac{\partial \kappa_j}{\partial u_{ik}}\mathrm{Im}(E_j H_j^*)\}, \quad (8)$$

where $\delta_{ik}$ is the Kronecker delta function. We call Eq. (8) the extended Helmholtz stress tensor because it is an extended form of traditional Helmholtz stress tensor [31, 36] that only works for achiral medium. Equation (8) is derived by applying virtual work principle based on free energy formulation of the EM problem. Therefore, it requires that the EM energy density, i.e. $-1/4\mathrm{Re}(\mathbf{E}_e \cdot \mathbf{D}_e^* + \mathbf{H}_e \cdot \mathbf{B}_e^*)$, of macroscopic effective medium must be equal to that of the microscopic metamaterial lattice. We proved in the Supplementary Materials Sec. VI [26] that such a relationship holds in the long-wavelength limit.

In the following, using examples with real structures, we will show that the propagating fields as well as the optical force density inside the bi-anisotropic metamaterials can be obtained using the Eqs. (6)-(8).

Consider the configuration shown in Fig 1(a), with a plane wave incident obliquely on the bi-anisotropic metamaterial slab with a square lattice structure, which has *N* layers along the *x* direction and is periodic along the *y* direction. In the long-wavelength limit, the metamaterial can be treated as an effective homogenous medium. The metamaterial, which is sandwiched by two layers of effective medium of the same type, should correspond to the effective medium slab shown in Fig. 1(b). We note that introducing the two layers of effective medium in Fig. 1(a) helps to reduce the boundary effect, which does not affect the physics discussed here. In the following, we consider two types of helical structures (labelled as type I and II) as shown in Fig. 1(c). The type I cylinder consists of a chain of helices with axis along the *z* direction, while the type II consists of a chain of helices with orthogonal axis along *x* and *y* directions, respectively. The helix chains are along z direction and have the same period $D$. Each helix has minor radius *r*, major radius *R* and contains 6 pitches with pitch length *p*. The helices are made of gold whose relative permittivity is described by the Drude model $\varepsilon_{Au} = 1 - \omega_p^2 / (\omega^2 + i\omega\omega_\tau)$ with plasma frequency $\omega_p = 1.36 \times 10^{16} s^{-1}$ and damping frequency $\omega_\tau = 4.084 \times 10^{13} s^{-1}$. As the pitch length *p* is much smaller than the major radius *R*, the helix is almost rotational invariant about its axis. As such, the Mie coefficients of the type I and II cylinders fulfill $A_n \approx A_{-n}, B_n \approx B_{-n}, C_n \approx C_{-n}$ (verified by the numerical calculations) which is the sufficient condition that the effective parameters of the metamaterials composed of these cylinders can be calculated using Eq. (6).

We first consider bi-anisotropic metamaterial composing of an array of type I chiral cylinders, where the helices are right-handed. The lattice constant of the metamaterial is set to be $a = 1\mu m$ which is much smaller than the wavelength $\lambda = 21.43\mu m$ of the incident wave (corresponding to the resonant frequency $f = 14\text{THz}$), and effective medium approaches should be good approximations in this regime. The quality of the effective constitutive parameters calculated by Eq. (6) can be checked by comparing the spatially averaged lattice fields $(1/\Omega\int_\Omega \mathbf{E}d\Omega, 1/\Omega\int_\Omega \mathbf{H}d\Omega)$ and the effective fields ($\mathbf{E}_e, \mathbf{H}_e$) in the corresponding effective medium [37]. With no loss of generality, we consider a $H_z$ polarized plane wave incident from the left hand side at an angle of $\theta = \pi/6$, and the comparison for the y component fields are shown in Fig 3(a). We can see that the spatially averaged lattice fields (circles) match the effective fields (lines) well. Such good agreement is also found for other field components, see Supplemental material Sec. V [26]. So the effective constitutive parameters provide an accurate description of the fields inside the metamaterial. In Supplemental material Sec. II. C [26], we also show the consistency between the numerical and analytical calculations of the electro/magneto-strictive tensors, indicating the validity of Eq. (7) for these structures.

To verify the correctness of the extended stress tensor, we did three types of calculations. For the first type, the total force acting on each chiral cylinder is calculated by integrating the Maxwell stress tensor over a boundary that encloses the cylinder [such as the dashed square in Fig. 1(a)]. The fields are obtained using full wave simulations. Dividing this force by the unit cell area gives the optical force density. The force density determined in this way is by definition correct because the cylinders are immersed in air. For the second type calculation, we integrate the extended Helmholtz stress tensor over the boundary that encloses the same region in the corresponding effective medium [such as the dashed square in Fig. 1(b)], using macroscopic effective fields and effective parameters. The obtained force divided by the same area corresponds to the optical force density within the effective medium framework. For the third type, we redid the effective medium calculation using the Maxwell stress tensor instead of the extended Helmholtz stress tensor. The results produced by the three different approaches are summarized in Fig. 3(b) as blue, red and black symbol-lines, respectively. We see that the extended Helmholtz stress tensor produces results that agree well with the microscopic lattice results, while the Maxwell stress tensor fails to do so.

We next consider the bi-anisotropic metamaterial composed of type II cylinders and assume the same lattice constant and incident wave as in type I. Figure 4(a) shows that the spatially averaged lattice fields and the effective fields in the corresponding effective medium are consistent with each other, indicating that the effective constitutive parameters are correctly obtained. Figure 4(b) shows the optical force density inside the metamaterial obtained using the Maxwell stress tensor under the full wave simulations and the optical force densities inside the corresponding effective medium obtained using both the extended Helmholtz and Maxwell stress tensors within effective medium formulism. We see that the extended Helmholtz stress tensor produces accurate optical force density. Note that in this case the Maxwell stress tensor within effective medium formulism produces almost the same result. This is because the effective permittivity and permeability $|\varepsilon_t|, |\mu_t|$ are very close to 1 so that the extended Helmholtz and Maxwell stress tensors produce are almost identical results. When $|\varepsilon_t|, |\mu_t|$ deviate from 1, for example in the case that the helices have a high dielectric cylinder core with relative permittivity $\varepsilon_c = 12.5$ as shown in the inset of Fig. 4(c), the Maxwell stress tensor will no longer give the correct description of optical force density. In Fig. 4(c), we show the consistency between the spatially averaged lattice fields and the fields inside the effective medium at the resonant frequency $f = 10.6$ THz. The comparison of the optical force densities calculated using different approaches is shown in Fig. 4(d). The Maxwell stress tensor gives the correct trend but not the magnitude of the optical force density while the results of the extended Helmholtz stress tensor are in accordance with that of the microscopic lattice.

In summary, using multiple scattering theory, we derived closed-form expressions for effective constitutive parameters and electro/magneto-strictive tensors components for 2D bi-anisotropic metamaterials. We also derived an expression for the extended electromagnetic stress tensor for these materials. The effective constitutive parameters can describe the optical scattering and absorption properties of bi-anisotropic metamaterials while the electro/magneto-strictive components, together with the extended electromagnetic stress tensor, provides sufficient information to determine the total optical force and the optical force density induced by external EM waves. We can use these macroscopic parameters to describe and predict the optical and opto-mechanical responses of complex man-made bi-anisotropic media, without the need to worry about the complex underlying structure. The results can also deepen our understanding of light-induced forces inside a complex medium and may find applications in optical manipulations, such as the optical stretching, compressing and sorting of materials.

**Acknowledgements**

We thank Dr. Kun Ding, Ruo-Yang Zhang and Xulin Zhang for valuable comments and suggestions. This work was supported by Hong Kong Research Grant Council grant AoE/P-02/12.

**References**

1. R. A. Shelby, D. R. smith, and S. Schultz, *Science* **292**, 77 (2001).
2. J. B. Pendry, *Phys. Rev. Lett.* **85**, 3966 (2000).
3. J. B. Pendry, D. Schurig, and D. R. Smith, *Science* **312**, 1780 (2006).
4. D. Schurig, J. J. Mock, B. J. Justice, S. A. Cummer, J. B. Pendry. A. F. Starr, and D. R. Smith, *Science* **314**, 977 (2006).
5. W. Cai, U. K. Chettiar, A. V. Kildishev, and V. M. Shalaev, *Nat. Photon.* **1**, 224 (2007).
6. H. Chen and C. T. Chan, *Appl. Phys. Lett*. **91**, 183518 (2007).
7. M. Silveirinha and N. Engheta, *Phys. Rev. Lett*. 97, 157403 (2006).
8. X. Huang, Y. Lai, Z. H. Hang, H. Zheng, and C. T. Chan, *Nat. Mater*. **10**, 582 (2011).
9. A. Priou, A. Sihvola, S. Tretyakow, *Advances in complex electromagnetic materials* (Springer Science & Business Media, 2012).
10. A. h. Sihvola and Olli P. M. Pekonen, J. Phys. D: App. Phys. **29**, 514 (1996).
11. G. W. Milton, The theory of composites, (Cambridge University Press 2002).
12. J. B. Pendry, *Science* **306**, 1353 (2004).
13. S. Zhang, Y. S. Park, J. Li, X. Lu, W. Zhang and X. Zhang, *Phys. Rev. Lett.* **102**, 023901 (2009).
14. C. Wu, H. Li, Z. Wei, X. Yu and C. T. Chan, *Phys. Rev. Lett.* **105**, 247401 (2010).
15. E. Plum, V. A. Fedotov, A. S. Schwanecke and N. I. Zheludev, *Appl. Phys. Let*. **90**, 223113 (2007).
16. H. Liu, D. A. Genov, D. M. Wu, Y. M. Liu, Z. W. Liu, C. Sun, S. N. Zhu and X. Zhang, *Phys. Rev. B* **76**, 073101 (2007).
17. T. Q. Li, H. Liu, T. Li, S. M. Wang, F. M. Wang, R. X, Wu, P. Chen, S. N. Zhu and X. Zhang, *Appl. Phys. Lett.* **92**, 131111(2008).
18. S. L. Prosvirnin and N. I. Zheludev, *Phys. Rev. E* **71**, 037603 (2005).
19. V. A. Fedotov, P. L. Mladyonov, S. L. Prosvirnin, A. V. Rogacheva, Y. Chen and N. I. Zheludev, *Phys. Rev. Lett*. **97**, 167401 (2006).
20. E. Plum, X.-X. Liu, V. A. Fedotov, Y. Chen, D. P. Tsai and N. I. Zheludev, *Phys. Rev. Lett*. **102**, 113902 (2009).
21. A. B. Khanikaev, S. H. Mousavi, W.-K. Tse, M. Kargarian, A. H. MacDonald and G. Shevets, *Nat. Mat.* **12**, 233 (2013).

22. W. Gao, M. Lawrence, B. Yang, F. Liu, F. Fang, B. Beri, J. Li and S. Zhang, *Phys. Rev. Lett.* **114**, 037402 (2015).
23. J. Guck, R. Ananthakrishnan, T. J. Moon, C. C. Cunningham, and J. Kas, *Phys. Rev. Lett*. **84**, 5451 (2000).
24. J Guck, R. Ananthakrishnan, H. Mahmood, T. J. Moon, C. C. Cunningham, and J. Kas, *Biophys. J*. **81**, 767 (2001).
25. B. F. Kennedy, P. Wijesinghe, and D. D. Sampson, *Nature Photon*. **11**, 215 (2017).
26. Supplemental material for detailed derivations of the effective medium theory, the extended Helmholtz stress tensor and the equality relationship between energy densities. It also includes the method for numerically calculating the Mie coefficients and the electro/magneto-strictive tensors.
27. Y. Wu and Z. Q. Zhang, *Phys. Rev. B* **79**, 195111 (2009).
28. W. Sun, S. B. Wang, J. Ng, L. Zhou, and C. T. Chan, *Phys. Rev. B* **91**, 235439 (2015).
29. A. H. Sihvola and I. V. Lindell, *Electron. Lett*. **26**, 118 (1990).
30. A. Lakhtakia, V. K. Varadan and V. V. Varadan, *J. Mater. Res.* **8**, 917 (1992).
31. L. D. Landau, E. M. Lifshitz, and L. P. Pitaevskii, *Electrodynamics of Continuous Media*, 2nd ed. (Butterworth-Heinemann, New York, 1984).
32. P. Penfield, and H. A. Haus, *Electrodynamics of Moving Media* (MIT Press, Cambridge, 1967).
33. L. D. Landau and E. M. Lifshitz, *Theory of Elasticity*, 3rd ed. (Butterworth-Heinemann, New York, 1986).
34. R. A. Anderson, *Phys. Rev. B* **33**, 1302 (1986).
35. Y. M. Shkel, and D. J. Klingenberg, *J. Appl. Phys.* **80**, 4566 (1996).
36. H. Helmholtz, *Ann. Phys.* **249**, 385 (1881).
37. M. G. Silveirinha, *Phys. Rev. B* **75**, 115104, (2007).

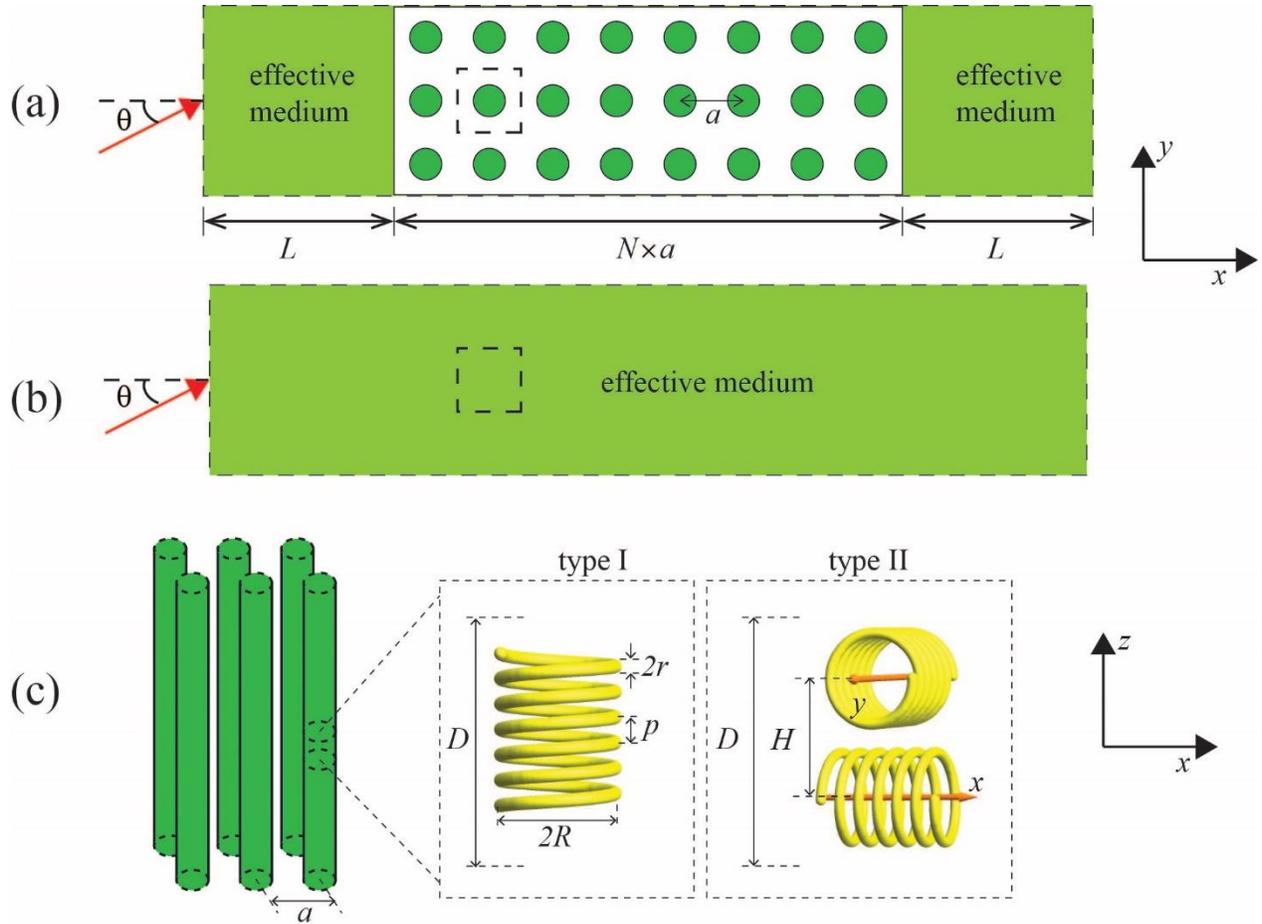

**Figure 1**. (a) The metamaterial slab is illuminated by a plane wave with incident angle $\theta$. (b) The corresponding effective homogenous medium. (c) The artificial chiral cylinders consist of identical gold helixes arranged into a column along z direction. Type I structure has chirality component $\kappa_z$ while type II contributes to a chirality component $\kappa_t$. The period of the unit cell along z direction is $D = 2H = 1.2$ μm. Each helix contains 6 pitches and we set $r = 15$ nm, $R = 200$ nm, $p = 100$ nm.

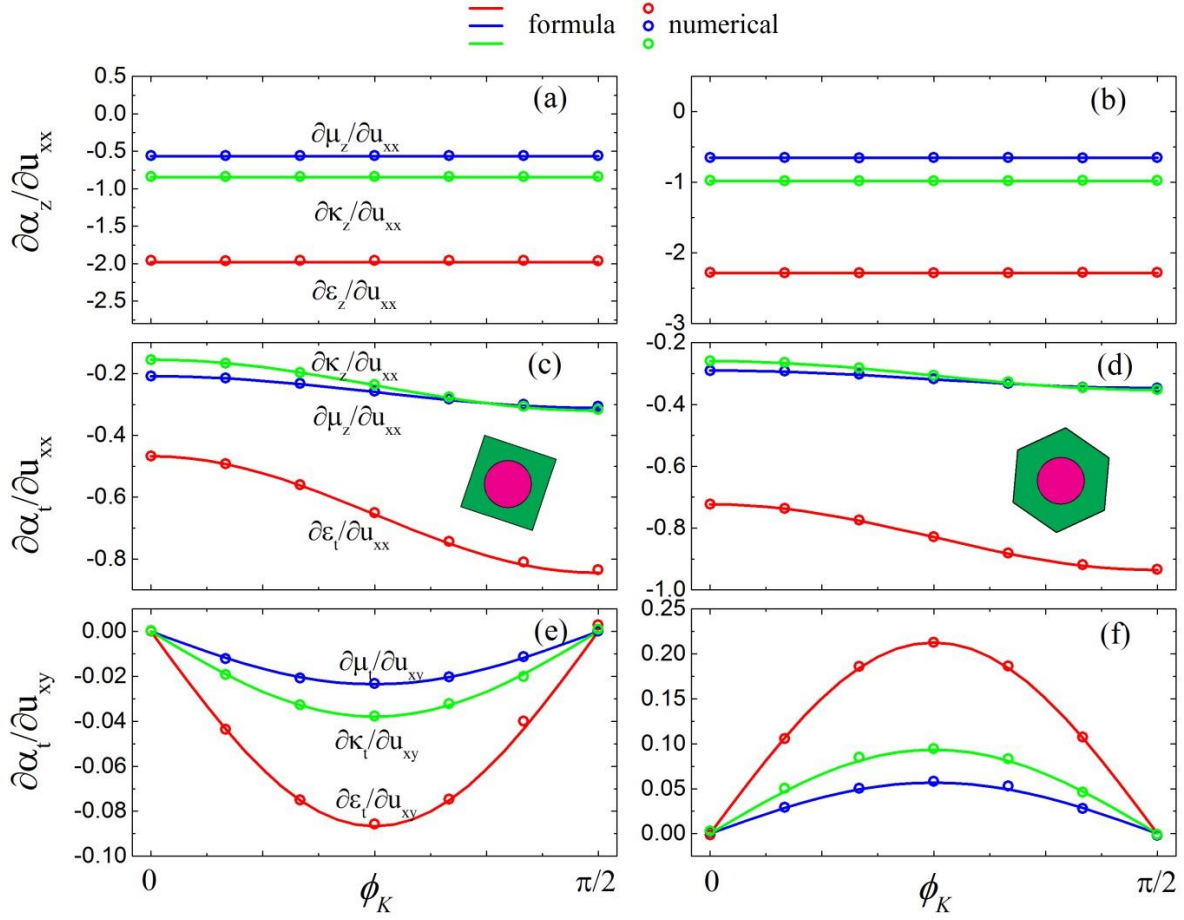

**Fig. 2.** Comparison between tensor components obtained from analytic formulas and those from numerical calculations for both square and hexagonal lattice structures. Results as functions of the direction of Bloch vector $\phi_K$ obtained from the formulas and numerical calculations are shown by lines and circles, respectively. $\alpha$ denotes $\varepsilon, \mu$ or $\kappa$. The inclusions of the metamaterials possess $\varepsilon = 8, \mu = 3, \kappa = 3, r_0 = 0.3a$.

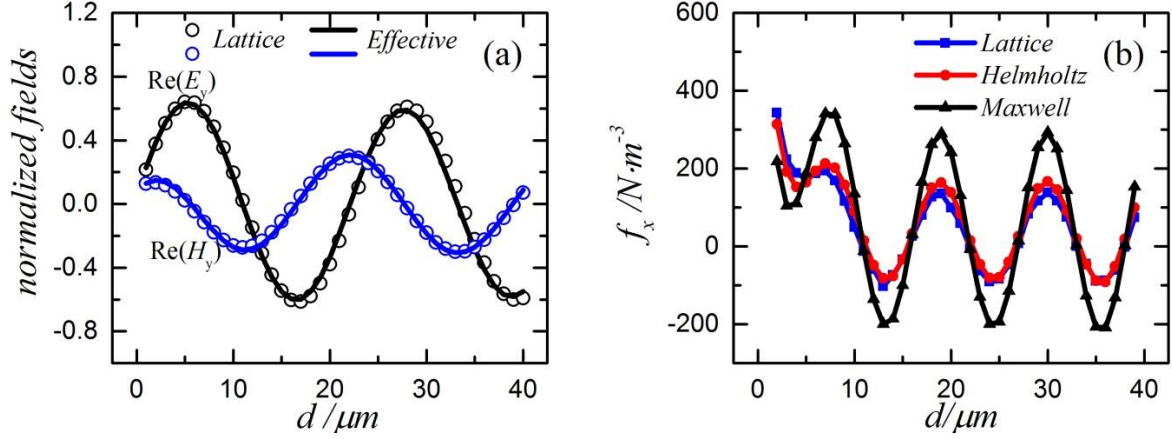

**Fig. 3.** (a) Comparison of the spatially averaged EM fields inside the metamaterial using full wave simulations (circles) and the effective fields in corresponding effective medium (lines) for type I structure. The field intensity is normalized by the incident fields. (b) Optical force densities calculated using different approaches. The incident plane wave is $H_z$ polarized with an amplitude of $E_0 = 10^5 V/m$. Other parameters are $N = 40, a = 1\mu m, L = 0.2\mu m, \theta = \pi/6, \lambda = 21.43\mu m$. The effective constitutive parameters for the metamaterials are given by Eq. (6) as $\varepsilon_z = 0.666 + 1.05i$, $\mu_z = 0.928 + 0.140i$, $\kappa_z = -0.160 + 0.392i$, $\varepsilon_t = 1.21$, $\mu_t = 1.00$, $\kappa_t = 0$.

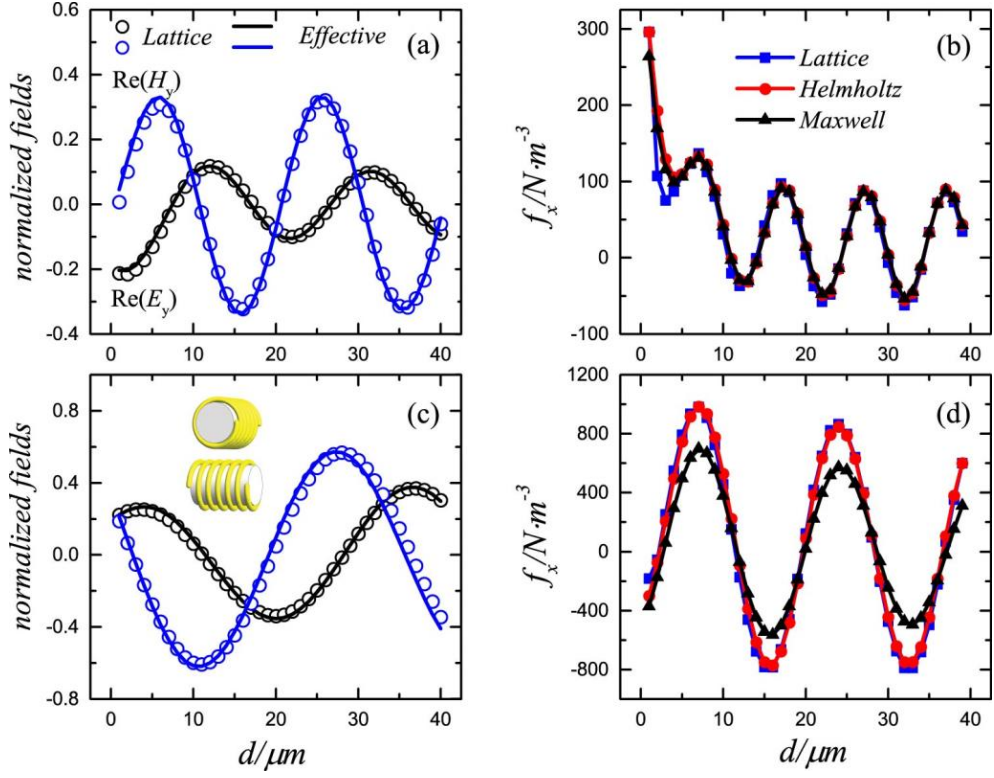

**Fig.4.** (a) Comparison of the spatially averaged EM fields inside the metamaterial (circles) and the effective fields in corresponding effective medium (lines) for type II cylinders. (b) Optical force density evaluated using different approaches. (c) Comparison of the spatially averaged EM fields inside the metamaterial formed of helices with a high dielectric ($\varepsilon_c = 12.5$) cylinder core. The dielectric core (shown by the inset) has height $H' = 0.6$ μm and radius $R' = 170$ nm. (d) Optical force density evaluated using different approaches for the metamaterial formed of helices with dielectric core. For panels (a) and (b), the incident wave is $H_z$ polarized with $\lambda = 21.43$ μm and $\theta = \pi/6$, and the effective parameters are $\varepsilon_z = 1.47, \mu_z = 1.00, \kappa_z = 0$, $\varepsilon_t = 0.751 + 0.835i, \mu_t = 0.933 + 0.129i, \kappa_t = -0.173 + 0.327i$. For panels (c) and (d), the incident wave is $E_z$ polarized with $\lambda = 28.28$ μm and $\theta = \pi/3$, and the effective parameters are $\varepsilon_z = 1.49, \mu_z = 1.0, \kappa_z = 0, \varepsilon_t = 1.26 + 0.42i, \mu_t = 0.969 + 0.116i, \kappa_t = -0.062 + 0.220i$. Other parameters are $N = 40, a = 1$ μm, $L = 1$ μm, $E_0 = 10^5$ V/m.

# Supplemental Material for "Closed-form expressions for effective constitutive parameters and electro/magneto-strictive tensors for bi-anisotropic metamaterials and their use in optical force density calculations"

Neng Wang[1], Shubo Wang[1,2], Zhao-Qing Zhang[1], and C. T. Chan[1*]

[1]*Department of Physics and Institute for Advanced Study, The Hong Kong University of Science and Technology, Hong Kong, China.*

[2]*Department of Physics, City University of Hong Kong, Hong Kong, China.*

Correspondence to: phchan@ust.hk

**Table of contents**

## Section I: Analytic derivations of the effective constitutive parameters and electro/magneto-strictive tensors

Using multiple scattering theory (MST), we derived closed-form expressions for the effective constitutive parameters as well as electro/magneto-strictive tensors for bi-anisotropic metamaterials in the long-wavelength limit. The former can be regarded as an extension of the traditional Maxwell-Garnet formula to bi-anisotropic metamaterials and the latter are useful for calculating the optical force distribution inside metamaterials.

### A. Mie theory for isotropic chiral cylinders

Consider the scattering of electromagnetic wave by a single chiral cylinder. The cylinder is isotropic with the constitutive relations given by

$$\mathbf{D} = \varepsilon\varepsilon_0 \mathbf{E} + i\kappa/c\mathbf{H}, \qquad \mathbf{B} = \mu\mu_0 \mathbf{H} - i\kappa/c\mathbf{E}.$$

Using the Mie scattering method [1], the incident field, scattered field and the field inside the chiral cylinder at a location $\mathbf{r} = (r, \phi)$ can be written as

$$\begin{aligned}
\mathbf{E}_i &= \sum_n [q_n \mathbf{N}_n^{(1)}(k_0, \mathbf{r}) + p_n \mathbf{M}_n^{(1)}(k_0, \mathbf{r})], \\
\mathbf{H}_i &= -i\sqrt{\frac{\varepsilon_0}{\mu_0}} \sum_n [p_n \mathbf{N}_n^{(1)}(k_0, \mathbf{r}) + q_n \mathbf{M}_n^{(1)}(k_0, \mathbf{r})], \\
\mathbf{E}_s &= -\sum_n [b_n \mathbf{N}_n^{(3)}(k_0, \mathbf{r}) + a_n \mathbf{M}_n^{(3)}(k_0, \mathbf{r})], \\
\mathbf{H}_s &= i\sqrt{\frac{\varepsilon_0}{\mu_0}} \sum_n [a_n \mathbf{N}_n^{(3)}(k_0, \mathbf{r}) + b_n \mathbf{M}_n^{(3)}(k_0, \mathbf{r})], \qquad (S1) \\
\mathbf{E}_{ins} &= \sum_n [c_n \mathbf{N}_n^{(1)}(k_1, \mathbf{r}) + c_n \mathbf{M}_n^{(1)}(k_1, \mathbf{r}) + d_n \mathbf{N}_n^{(1)}(k_2, \mathbf{r}) - d_n \mathbf{M}_n^{(1)}(k_2, \mathbf{r})], \\
\mathbf{H}_{ins} &= -i\sqrt{\frac{\varepsilon_0 \varepsilon}{\mu_0 \mu}} \sum_n [c_n \mathbf{N}_n^{(1)}(k_1, \mathbf{r}) + c_n \mathbf{M}_n^{(1)}(k_1, \mathbf{r}) - d_n \mathbf{N}_n^{(1)}(k_2, \mathbf{r}) + d_n \mathbf{M}_n^{(1)}(k_2, \mathbf{r})],
\end{aligned}$$

where $k_0 = \omega/c$ is the wavenumber in the background, $k_{1,2} = (\sqrt{\varepsilon\mu} \pm \kappa)k_0$ are wavenumbers inside the chiral cylinder, and $\mathbf{M}_n^{(J)}, \mathbf{N}_n^{(J)}$ are the vector cylindrical wave functions [1] which are expressed as

$$\mathbf{M}_n^{(J)}(k,\mathbf{r}) = [\frac{in}{kr} z_n^{(J)}(kr)\mathbf{e}_r - z_n^{(J)}{}'(kr)\mathbf{e}_\phi]e^{in\phi},$$

$$\mathbf{N}_n^{(J)}(k,\mathbf{r}) = z_n^{(J)}(kr)e^{in\phi}\mathbf{e}_z.$$

with $z_n^{(1)}(kr) = J_n(kr), z_n^{(1)}{}'(kr) = J_n{}'(kr)$ denoting the Bessel function and its derivative with respect to its argument, and $z_n^{(3)}(kr) = H_n^{(1)}(kr), z_n^{(3)}{}'(kr) = H_n^{(1)}{}'(kr)$ denoting the Hankel function of first kind and its derivative with respect to its argument. The expansion coefficients for the incident and scattered fields are related by the Mie coefficients as $a_n = A_n p_n + C_n q_n$, $b_n = C_n p_n + B_n q_n$. [1] Employing the boundary conditions that the tangential electromagnetic fields should be continuous, we have

$$q_n J_n(x_0) - b_n H_n^{(1)}(x_0) = c_n J_n(x_1) + d_n J_n(x_2),$$
$$p_n J_n{}'(x_0) - a_n H_n^{(1)}{}'(x_0) = c_n J_n{}'(x_1) - d_n J_n{}'(x_2), \qquad (S2)$$
$$p_n J_n(x_0) - a_n H_n^{(1)}(x_0) = \sqrt{\frac{\varepsilon}{\mu}}[c_n J_n(x_1) - d_n J_n(x_2)],$$
$$q_n J_n{}'(x_0) - b_n H_n^{(1)}{}'(x_0) = \sqrt{\frac{\varepsilon}{\mu}}[c_n J_n{}'(x_1) + d_n J_n{}'(x_2)].$$

where $x_{0,1,2} = k_{0,1,2} r_0$ are dimensionless size parameters. The above equations can be solved to obtain the expressions for the Mie coefficients. In the long-wavelength limit, the zeroth and first-order Mie coefficients are reduced to

$$A_0 = \frac{i}{4}(1-\mu)\pi x_0^2, \quad A_1 = -\frac{i}{4}\frac{1+\varepsilon-\mu+\varepsilon\mu-\kappa^2}{1+\varepsilon+\mu+\varepsilon\mu-\kappa^2}\pi x_0^2,$$

$$B_0 = \frac{i}{4}(1-\varepsilon)\pi x_0^2, \quad B_1 = -\frac{i}{4}\frac{1-\varepsilon+\mu+\varepsilon\mu-\kappa^2}{1+\varepsilon+\mu+\varepsilon\mu-\kappa^2}\pi x_0^2, \qquad (S3)$$

$$C_0 = -\frac{i}{4}\kappa\pi x_0^2, \quad C_1 = -\frac{i}{2}\frac{\kappa}{1+\varepsilon+\mu+\varepsilon\mu-\kappa^2}\pi x_0^2.$$

**B. Multiple scattering formulism for bi-anisotropic cylinders**

In this subsection, we will derive the secular equation starting from the MST. The chiral cylinders actually do not need to be isotropic, and their scattering properties can still be described by the Mie coefficients $A_n, B_n, C_n$ which can become very complicated and cannot be expressed in closed form as in S3. Here we focus on the case that $A_n = A_{-n}, B_n = B_{-n}, C_n = C_{-n}$, which is true when the constitutive parameters of the cylinders are given by Eq. (1) in the main text.

For a 2D periodic system with multiple scattering between the cylinders, the incident field acting on an arbitrary cylinder $j$ also includes the scattering fields from other cylinders, which can be written as

$$\mathbf{E}_i(j) = -\sum_{l \neq j}\sum_m [a_m^l \mathbf{M}_m^{(3)}(k_0, \mathbf{r}-\mathbf{r}_l) + b_m^l \mathbf{N}_m^{(3)}(k_0, \mathbf{r}-\mathbf{r}_l)],$$

$$\mathbf{H}_i(j) = i\sqrt{\frac{\varepsilon_0}{\mu_0}} \sum_{l \neq j}\sum_m [b_m^l \mathbf{M}_m^{(3)}(k_0, \mathbf{r}-\mathbf{r}_l) + a_m^l \mathbf{N}_m^{(3)}(k_0, \mathbf{r}-\mathbf{r}_l)], \quad (S4)$$

where $a_m^l, b_m^l$ and $\mathbf{r}_l$ are scattering coefficients and location of cylinder $l$. Using the translation additional theorem [2]

$$\mathbf{M}_m^{(3)}(k_0, \mathbf{r}-\mathbf{r}_l) = \sum_n H_{m-n}^{(1)}(kd_{lj}) e^{-i(n-m)\phi_{lj}} \mathbf{M}_n^{(1)}(k_0, \mathbf{r}-\mathbf{r}_j),$$

$$\mathbf{N}_m^{(3)}(k_0, \mathbf{r}-\mathbf{r}_l) = \sum_n H_{m-n}^{(1)}(kd_{lj}) e^{-i(n-m)\phi_{lj}} \mathbf{N}_n^{(1)}(k_0, \mathbf{r}-\mathbf{r}_j), \quad (S5)$$

where $\mathbf{r}_{lj} = (d_{lj}, \phi_{lj})$ is the vector that directs from cylinder $j$ to cylinder $l$ and imposing the Bloch condition

$$a_m^l = a_m^j e^{i\mathbf{K}\cdot\mathbf{r}_{lj}}, \quad b_m^l = b_m^j e^{i\mathbf{K}\cdot\mathbf{r}_{lj}},$$

where $\mathbf{K} = (K, \phi_K)$ is the Bloch vector, the incident fields can be rewritten as

$$\mathbf{E}_i(j) = -\sum_n [\sum_m a_m^j S_{m-n} \mathbf{M}_n^{(1)}(k_0, \mathbf{r}-\mathbf{r}_j) + \sum_m b_m^j S_{m-n} \mathbf{N}_n^{(1)}(k_0, \mathbf{r}-\mathbf{r}_j)],$$

$$\mathbf{H}_i(j) = i\sqrt{\frac{\varepsilon_0}{\mu_0}} \sum_n [\sum_m b_m^j S_{m-n} \mathbf{M}_n^{(1)}(k_0, \mathbf{r}-\mathbf{r}_j) + \sum_m a_m^j S_{m-n} \mathbf{N}_n^{(1)}(k_0, \mathbf{r}-\mathbf{r}_j)], \quad (S6)$$

where $S_{m-n}$ is the lattice sum defined as

$$S_{m-n} = \sum_{l \neq j} e^{i\mathbf{K} \cdot \mathbf{r}_{lj}} H_{m-n}^{(1)}(kd_{lj}) e^{-i(n-m)\phi_{lj}}, \quad S_{n-m} = -(S_{m-n})^*.$$

Then we have the self-consistent equations

$$a_n^j = A_n \sum_m a_m^j S_{m-n} + C_n \sum_m b_m^j S_{m-n},$$

$$b_n^j = C_n \sum_m a_m^j S_{m-n} + B_n \sum_m b_m^j S_{m-n}, \quad (S7)$$

The dispersion relation $\omega(\mathbf{K})$ is obtained from the condition that gives nontrivial solutions to Eq. (S7) [3]. Up to dipole orders, the secular equation is reduced to

$$\det \begin{vmatrix} A_1 S_0 + 1 & A_1 S_{-1} & A_1 S_{-2} & C_1 S_0 & C_1 S_1 & C_1 S_{-2} \\ A_0 S_1 & A_0 S_0 + 1 & A_0 S_{-1} & C_0 S_1 & C_0 S_0 & C_0 S_{-1} \\ A_1 S_2 & A_1 S_1 & A_1 S_0 + 1 & C_1 S_2 & C_1 S_1 & C_1 S_0 \\ C_1 S_0 & C_1 S_{-1} & C_1 S_{-2} & B_1 S_0 + 1 & B_1 S_{-1} & B_1 S_{-2} \\ C_0 S_1 & C_0 S_0 & C_0 S_{-1} & B_0 S_1 & B_0 S_0 + 1 & B_0 S_{-1} \\ C_1 S_2 & C_1 S_1 & C_1 S_0 & B_1 S_2 & B_1 S_1 & B_1 S_0 + 1 \end{vmatrix} = 0. \quad (S8)$$

In the long-wavelength limit where $\omega \to 0, K \to 0$, the lattice sum can be decoupled as [3, 4]

$$S_n = \frac{4i^{n+1}}{k_0^2 \Omega} \frac{K^n}{k_0^n (k_0^2 - K^2)} e^{-in\phi_K} - \frac{2^{n+3} i^{n+1}(n+1)!}{k_0^n a^{n-2} \Omega} \sum_{K_h \neq 0} \frac{J_{n+1}(K_h a)}{(K_h a)^3} e^{-in\phi_{K_h}}, \quad (S9)$$

where $a$ denotes the lattice constant, $\Omega$ is the volume of the unit cell, and $\mathbf{K}_h = (K_h, \phi_{K_h})$ is the reciprocal-lattice vector. Then in the long-wavelength limit, the lowest order lattice sums are expressed as [3, 5]

$$S_0 = \frac{4i}{k_0^2 \Omega} \frac{1}{k_0^2 - K^2}, \quad S_{\pm 1} = \mp \frac{4}{k_0^2 \Omega} \frac{K}{k_0(k_0^2 - K^2)} e^{\mp i\phi_K}, \quad S_{\pm 2} = -\frac{4i}{k_0^2 \Omega} \frac{K^2}{k_0^2(k_0^2 - K^2)} e^{\mp 2i\phi_K}. \quad (S10)$$

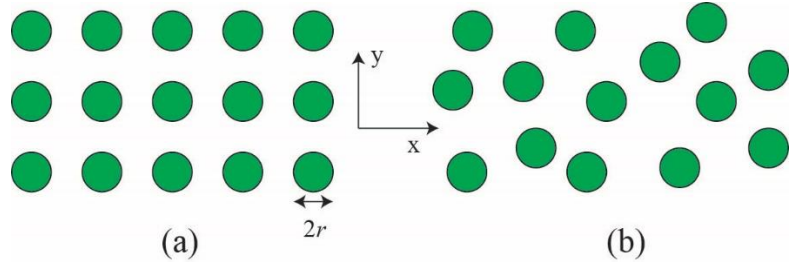

**Fig. S2**: top view of the bi-anisotropic metamaterial with parallel cylindrical inclusions arranged in either (a) regular or (b) random lattice in the $xy$ plane. The Mie coefficients of cylindrical inclusions satisfy $A_n = A_{-n}, B_n = B_{-n}, C_n = C_{-n}$ and the background is isotropic and achiral.

## C. Derivation of the effective constitutive parameters

In this section, we will obtain the closed-form expressions for the effective parameters for bi-anisotropic metamaterials composed of parallel inclusions arranged into a regular or random lattice, see Fiq. S1. The inclusions can be isotropic with Mie coefficients expressed as Eq. (S3) or anisotropic with Mie coefficients satisfy $A_n = A_{-n}, B_n = B_{-n}, C_n = C_{-n}$. Effective constitutive parameters of these metamaterials can be written as Eq. (1) and their expressions can be obtained by comparing the dispersion relations for the effective medium and the metamaterials.

For the effective medium, the electromagnetic wave has two eigen-modes with corresponding wave numbers expressed as

$$K_\pm^2 = k_0^2 \frac{\varepsilon_z \mu_t + \varepsilon_t \mu_z + 2\kappa_z \kappa_t \pm \sqrt{(\varepsilon_z \mu_t - \varepsilon_t \mu_z)^2 + 4\kappa_z \kappa_t (\varepsilon_z \mu_t + \varepsilon_t \mu_z) + 4\kappa_z^2 \varepsilon_t \mu_t + 4\kappa_t^2 \varepsilon_z \mu_z}}{2}, \qquad (S11)$$

which leads to the following relationships

$$K_+^2 K_-^2 = k_0^4 (\varepsilon_z \mu_z - \kappa_z^2)(\varepsilon_t \mu_t - \kappa_t^2). \qquad (S12)$$

The dispersion relation for a periodic system can be obtained by considering the secular equation. Substituting Eq. (S10) into Eq. (S8), the secular equation reduces to

$$K_+^2 K_-^2 = k_0^4 \frac{RS}{P}. \tag{S13}$$

where

$$\begin{aligned} P &= (i + \Lambda A_1)(i + \Lambda B_1) - \Lambda^2 C_1^2, \\ R &= (1 + i\Lambda A_0)(1 + i\Lambda B_0) + \Lambda^2 C_0^2, \\ S &= (i - \Lambda A_1)(i - \Lambda B_1) + \Lambda^2 C_1^2, \end{aligned} \tag{S14}$$

From Eq. (S14), we can see that $R$ and $S/P$ are related to the zeroth and first order Mie coefficients, respectively. And we note that the zeroth and first order Mie coefficients are related to the monopoles and dipoles which correspond to the $z$ and transverse components, respectively. Thus, comparing Eq. (S12) with Eq. (S13), we obtain

$$\varepsilon_z \mu_z - \kappa_z^2 = R, \quad \varepsilon_t \mu_t - \kappa_t^2 = \frac{S}{P}. \tag{S15}$$

Substitute Eq. (S14) into the above equation and we can obtain the out-plane components of the effective parameters as

$$\varepsilon_z = 1 + i\Lambda B_0, \quad \mu_z = 1 + i\Lambda A_0, \quad \kappa_z = i\Lambda C_0. \tag{S16a}$$

The expressions for in-plane components of the effective parameters can be determined with the following considerations: $\varepsilon_t$ and $\mu_t$ should be interchanged when $A_1$ and $B_1$ are interchanged; they are even functions of $C_1$; they can be reduced to the achiral forms of

$$\varepsilon_t = \frac{(i - \Lambda A_1)}{(i + \Lambda A_1)}, \qquad \mu_t = \frac{(i + \Lambda A_1)}{(i + \Lambda A_1)}, \tag{S17}$$

when $C_1$ is zero; for $\kappa_t$, it is an odd function of $C_1$. With all this information, we can obtain

$$\varepsilon_t = \frac{(i-\Lambda A_1)(i+\Lambda B_1)+\Lambda^2 C_1^2}{(i+\Lambda A_1)(i+\Lambda B_1)-\Lambda^2 C_1^2},$$

$$\mu_t = \frac{(i+\Lambda A_1)(i-\Lambda B_1)+\Lambda^2 C_1^2}{(i+\Lambda A_1)(i+\Lambda B_1)-\Lambda^2 C_1^2}, \quad \text{(S16b)}$$

$$\kappa_t = \frac{-2i\Lambda C_1}{(i+\Lambda A_1)(i+\Lambda B_1)-\Lambda^2 C_1^2}.$$

For special case of isotropic inclusions, substituting Eq. (S3) into the above equations, then we have

$$\varepsilon_z = (\varepsilon-1)p+1, \quad \mu_z = (\mu-1)p+1, \quad \kappa_z = \kappa p, \quad \text{(S18a)}$$

and

$$\varepsilon_t = \frac{(\varepsilon+1)(\mu+1)-\kappa^2+2p(\varepsilon-\mu)-p^2((\varepsilon-1)(\mu-1)-\kappa^2)}{(\varepsilon+1)(\mu+1)-\kappa^2+2p(1+\kappa^2-\varepsilon\mu)+p^2[(\varepsilon-1)(\mu-1)-\kappa^2]},$$

$$\mu_t = \frac{(\varepsilon+1)(\mu+1)-\kappa^2-2p(\varepsilon-\mu)-p^2((\varepsilon-1)(\mu-1)-\kappa^2)}{(\varepsilon+1)(\mu+1)-\kappa^2+2p(1+\kappa^2-\varepsilon\mu)+p^2[(\varepsilon-1)(\mu-1)-\kappa^2]}, \quad \text{(S18b)}$$

$$\kappa_t = \frac{4p\kappa}{(\varepsilon+1)(\mu+1)-\kappa^2+2p(1+\kappa^2-\varepsilon\mu)+p^2[(\varepsilon-1)(\mu-1)-\kappa^2]},$$

where $p = \pi x_0^2/(k_0^2\Omega)$ denotes the filling ratio with $x_0 = k_0 r_0$ being the dimensionless size parameter of the inclusions. We note that Eq. (S17) is more fundamental than Eq. (S18) since it does not require the inclusions to be isotropic. For anisotropic inclusions with constitutive parameters such as Eq. (1), the effective constitutive parameters can be obtained using Eq. (S17) once the Mie coefficients of the inclusions are known. Note that Eqs. (S17) and (S18) are valid for both regular and random lattice structures in the long-wavelength limit, since they are only related to the scattering properties of inclusions and the filling ratios. And when $\kappa = 0$, Eq. (S18) reduces to the well-known traditional 2D Maxwell Garnett formula.

From Eq. (S18), we can obtain the following relationship

$$\frac{\varepsilon_z - \mu_z}{\kappa_z} = \frac{\varepsilon_t - \mu_t}{\kappa_t} = \frac{\varepsilon - \mu}{\kappa}. \quad \text{(S19)}$$

The same relationship has been found in 3D isotropic chiral metamaterials [6]. The identity (S19) indicates that the relation among effective constitutive parameters of the metamaterial is essentially equal to that of the inclusions. Therefore we can tune the effective constitutive parameters by adjusting the constitutive parameters of the inclusions according to Eq. (S19).

Also, combining Eq. (S18b) and the Eq. (9) in Refs. [6], a general expression for both the 2D and 3D chiral metamaterials is found as

$$\varepsilon_t = \frac{[\alpha+\varepsilon-p(1-\varepsilon)][\alpha+\mu+p(1-\mu)]+\kappa^2 p^2(p-1)(1+\alpha p)}{[\alpha+\varepsilon+p(1-\varepsilon)][\alpha+\mu+p(1-\mu)]-\kappa^2(p-1)^2},$$

$$\mu_t = \frac{[\alpha+\varepsilon+p(1-\varepsilon)][\alpha+\mu-p(1-\mu)]+\kappa^2 p^2(p-1)(1+\alpha p)}{[\alpha+\varepsilon+p(1-\varepsilon)][\alpha+\mu+p(1-\mu)]-\kappa^2(p-1)^2}, \quad (S20)$$

$$\kappa_t = \frac{(\alpha+1)^2 p\kappa}{[\alpha+\varepsilon+p(1-\varepsilon)][\alpha+\mu+p(1-\mu)]-\kappa^2(p-1)^2},$$

where $\alpha = d-1$ with $d$ being the dimension of the system.

In the following, we will numerically check the validity of Eq. (S18). We consider a square/hexagonal lattice formed by 50 layers of cylinders ($\varepsilon=8, \mu=1, \kappa=2$) in the $x$ direction and is periodic along the $y$ direction. When the lattice constant $a$ is much smaller than the wavelength, such a lattice can be regarded as an effective slap according to the EMT and the corresponding effective constitutive parameters can be obtained using Eq. (S18). The validity of the effective parameters can be tested by checking the consistence between the spatially averaged lattice fields, such as the electric field and displacement field: $\bar{\mathbf{E}} = 1/\Omega \int_\Omega \mathbf{E} dS$, $\bar{\mathbf{D}} = 1/\Omega \int_\Omega \mathbf{D} dS$, and the fields in the corresponding effective medium [7]. To do this, we consider a $H_z$ polarized plane wave obliquely incident on the lattices, as shown in the insets of Fig.1, and calculate the spatially averaged lattice fields using a commercial finite-element-method package COMSOL [8]. The electric field and displacement field along the $x$ direction inside the corresponding effective medium are also numerically computed. We can see that for both square [Fig.S2(a)] and hexagonal lattices [Fig.S2(b)], the spatially averaged lattice fields (symbols) are in accordance with the fields in effective mediums (lines). Such consistence also exists for the magnetic field. This indicates that our formulas Eq. (S18) can correctly determine the effective constitutive parameters for this kind of bi-anisotropic metamaterials.

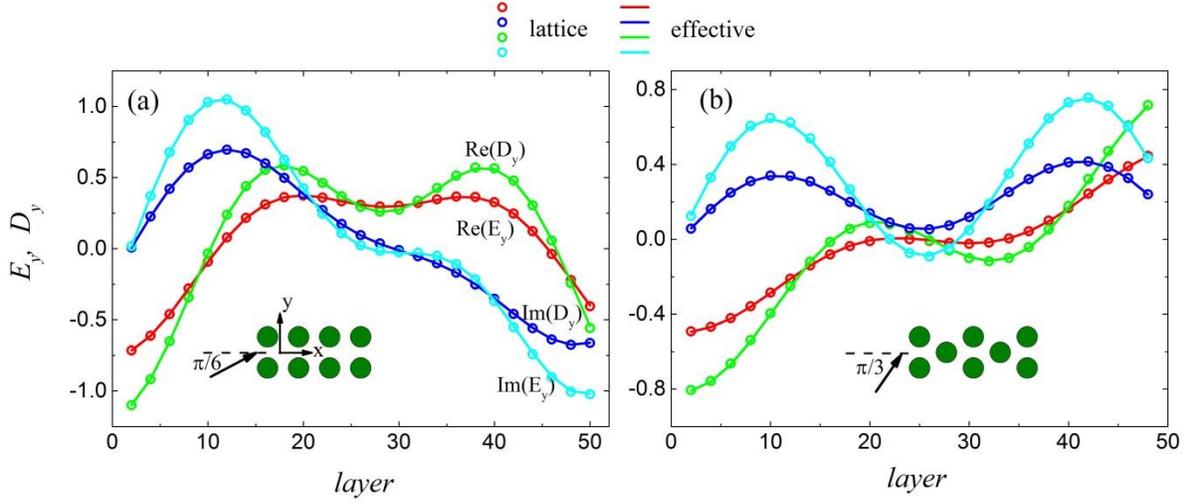

**Fig. S2.** The spatially averaged electric field and displacement field of each layer inside the metamaterials with (a) the square lattice and (b) hexagonal lattice are noted by the circles. The electric field and displacement field along the $x$ direction in each corresponding effective continuous medium under same incidence are shown by lines. The inclusions of the metamaterials possess $\varepsilon = 8, \mu = 1, \kappa = 2, r_0 = 0.3a$.

## D. Derivation of the electro/magneto-strictive tensors

The electro/magneto-strictive tensors are defined as $\partial \varepsilon / \partial u_{ik}, \partial \mu / \partial u_{ik}$ and $\partial \kappa / \partial u_{ik}$ for the permittivity, permeability and chirality parts, respectively. Here $u_{ik} = (\partial u_i / \partial x_k + \partial u_k / \partial x_i)/2$ is the strain tensor with $\mathbf{u}(x)$ being the displacement vector [9]. The electrostrictive and magnetostrictive tensors describe the "stiffness" of constitutive parameters under stretching (diagonal terms in $u_{ik}$) and shearing (off diagonal terms in $u_{ik}$) [10,11]. In Fig. 3, the geometrical sketches of the stretched and sheared unit cells for square and hexagonal lattices are shown, and the strain tensor is given by $u_{xx} = 2\Delta a / a, u_{xy} = \Delta a / a$ for the square lattice and $u_{xx} = 2\Delta a / a, u_{xy} = \Delta a / (\sqrt{3}a)$ for hexagonal lattice, where $\Delta a$ is an infinitesimal displacement.

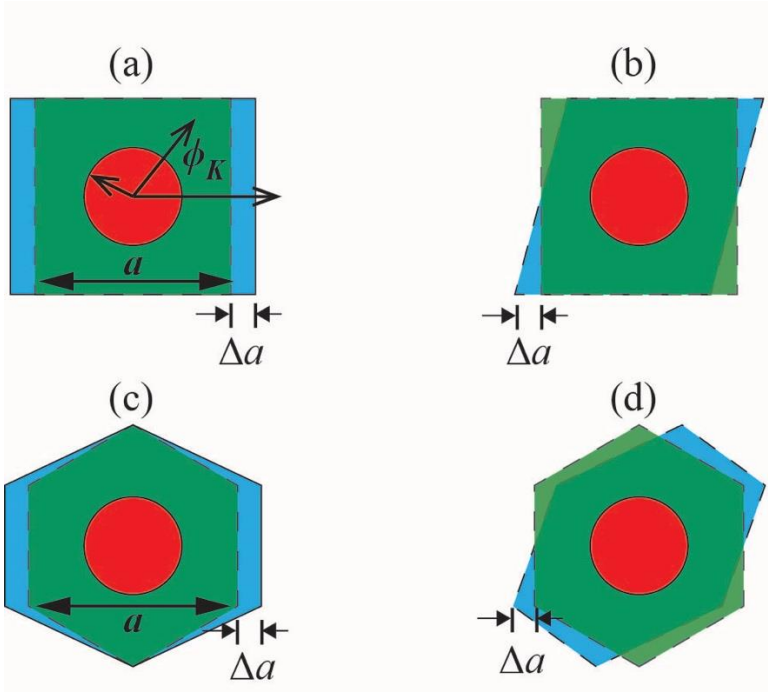

**Fig. S3.** The unit cell deformations used to calculate the electrostrictive and magnetostrictive tensors. The upper (lower) panels are for the square (hexagonal) lattice. (a), (b) Original square lattice unit cell is shown as semitransparent green squares. The unit cell is stretched or sheared $\Delta a$ in $x$ direction, with the deformed cell shown by blue parallelograms. Here $a$ and $\phi_K$ denote the side length of cell and the direction of wave vector, respectively. (c), (d) Counterparts for the hexagonal lattice.

For the lattice stretched along the $x$ direction, the lattice sums can be expressed as [5]

$$S_0 = i\Lambda \frac{1}{k_0^2 - K^2}(1 - u_{xx}),$$

$$S_{\pm 1} = \mp \Lambda \frac{K}{k_0(k_0^2 - K^2)} e^{\mp i\phi_K}(1 - u_{xx}), \quad (S21)$$

$$S_{\pm 2} = (-i\Lambda \frac{K^2}{k_0^2(k_0^2 - K^2)} e^{\mp 2i\phi_K} + i\Pi u_{xx})(1 - u_{xx}),$$

where $\Pi = 1.298\Lambda$ for the square lattice and $\Pi = 0.499\Lambda$ for the hexagonal lattice [5]. Substituting Eq. (S21) into the secular equation Eq. (S8), and then we obtain $K_+^2 K_-^2$ as a function of $u_{xx}$. We thus have

$$\frac{1}{k_0^4}\frac{\partial (K_+^2 K_-^2)}{\partial u_{xx}}\bigg|_{u_{xx} \to 0} = \frac{\partial}{\partial u_{xx}}(\varepsilon_z \mu_z - \kappa_z^2)(\varepsilon_t \mu_t - \kappa_t^2) = [-\Lambda(A_0 + B_0) + 2\Lambda(A_0 B_0 - C_0^2)]\frac{S}{P} - R\frac{T}{P^2}, \quad (S22)$$

where $P, R, S$ are defined in Eq. (S14), and

$$T = 2\Lambda\{\Pi B_1^2 + A_1^2[\Pi + \Lambda^2 B_1(i + 2\Pi B_1)] - iB_1(-1 + \Lambda^2 C_1^2) + 2\Pi C_1^2(1 + \Lambda^2 C_1^2)$$
$$+ iA_1[1 + \Lambda^2(B_1^2 - C_1^2 + 4i\Pi B_1 C_1^2)]\}. \tag{S23}$$

We already know that $(\varepsilon_z \mu_z - \kappa_z^2) = R$, $(\varepsilon_t \mu_t - \kappa_t^2) = S/P$, see Eq. (S15). Thus we have

$$\frac{\partial}{\partial u_{xx}}(\varepsilon_z \mu_z - \kappa_z^2) = -\Lambda(A_0 + B_0) + 2\Lambda(A_0 B_0 - C_0^2), \tag{S24}$$

and

$$\frac{\partial}{\partial u_{xx}}(\varepsilon_t \mu_t - \kappa_t^2) = -\frac{T}{P^2}. \tag{S25}$$

According to Eq. (S24) and Eq. (S16a), we could easily obtain that

$$\frac{\partial \varepsilon_z}{\partial u_{xx}} = -i\Lambda B_0 = 1 - \varepsilon_z, \quad \frac{\partial \mu_z}{\partial u_{xx}} = -i\Lambda A_0 = 1 - \mu_z, \quad \frac{\partial \kappa_z}{\partial u_{xx}} = i\Lambda C_0 = -\kappa_z. \tag{S26}$$

For the in-plane components, we write

$$\frac{\partial \varepsilon_t}{\partial u_{xx}} = \frac{2\Lambda A_1(i + \Pi A_1)(i + \Lambda B_1)^2 - g_1 C_1^2 + 2\Lambda^3 \Pi C_1^4}{[(i + \Lambda A_1)(i + \Lambda B_1) - \Lambda^2 C_1^2]^2},$$

$$\frac{\partial \mu_t}{\partial u_{xx}} = \frac{2\Lambda B_1(i + \Pi B_1)(i + \Lambda A_1)^2 - g_2 C_1^2 + 2\Lambda^3 \Pi C_1^4}{[(i + \Lambda A_1)(i + \Lambda B_1) - \Lambda^2 C_1^2]^2}, \tag{S27}$$

$$\frac{\partial \kappa_t}{\partial u_{xx}} = \frac{f_1 C_1 + f_2 C_1^3}{[(i + \Lambda A_1)(i + \Lambda B_1) - \Lambda^2 C_1^2]^2},$$

where $g_1, g_2$ and $f_1, f_2$ are coefficients to be determined. We write $\partial \varepsilon_t / \partial u_{xx}$ and $\partial \mu_t / \partial u_{xx}$ in such forms by the consideration of three aspects. First, they should be interchanged when $A_1$ and $B_1$ are interchanged. Second, they should be even functions of $C_1$ (or $\kappa$). Third, they should be reduced to the achiral forms

$$\frac{\partial \varepsilon_t}{\partial u_{xx}} = \frac{2\Lambda A_1(i + \Pi A_1)}{(i + \Lambda A_1)^2}, \quad \frac{\partial \mu_t}{\partial u_{xx}} = \frac{2\Lambda B_1(i + \Pi B_1)}{(i + \Lambda A_1)^2}, \tag{S28}$$

when $C_1$ vanishes at $\kappa = 0$. For the chirality, $\partial \kappa_t / \partial u_{xx}$ should be an odd function of $C_1$.

Substituting Eqs. (S28) and (S16b) into Eq. (S25), we could have

$$\begin{aligned}
g_1 + g_2 &= -2\Lambda[-4\Lambda + i\Lambda(A_1^2 + B_1^2) + 2\Pi + 2i\Lambda\Pi(A_1 + B_1) + 4\Lambda^2\Pi A_1 B_1], \\
f_1 &= -\Lambda\{2i + 3\Pi A_1 + \Pi B_1 + \Lambda B_1[2\Pi A_1^2 - iB_1 + A(3i - 2\Pi B_1)] \\
&\quad + 2\Lambda[i\Pi A_1^2 + B_1 - A_1(1 + 3i\Pi B_1)]\} - \frac{1}{2}(A_1 - B_1)g_1, \\
f_2 &= -2i\Lambda(-\Lambda^2 + 2\Lambda\Pi).
\end{aligned} \tag{S29}$$

Up to now, $g_1, g_2$ and $f_1$ are still undetermined. We use another information that Eq. (S27) should be reduced to the form of the amorphous metamaterials when $\Pi = 0$. The electrostrictive and magnetostrictive tensors for amorphous metamaterials can be directly obtained according to Eq. (S16b), namely

$$\frac{\partial \varepsilon_t}{\partial u_{xx}} = -p\frac{\partial \varepsilon_t}{\partial p} = \frac{2\Lambda[iA_1(i + \Lambda B_1)^2 + \Lambda(2 - i\Lambda B_1)C_1^2]}{[(i + \Lambda A_1)(i + \Lambda B_1) - \Lambda^2 C_1^2]^2} = \frac{2\Lambda A_1 i(i + \Lambda B_1)^2 - g_1 C_1^2}{[(i + \Lambda A_1)(i + \Lambda B_1) - \Lambda^2 C_1^2]^2}\bigg|_{\Pi=0}. \tag{S30}$$

And considering that $g_1$ and $g_2$ should be interchanged when interchanging $A_1$ and $B_1$, we have

$$g_1 = -2\Lambda(-2\Lambda + i\Lambda^2 B_1 + \Pi + 2i\Lambda\Pi A_1 + 2\Lambda^2\Pi A_1 B_1). \tag{S31}$$

Substituting Eq. (S31) into Eq. (S29) and combining Eqs. (S27) and (S16b), we finally have

$$\begin{aligned}
\frac{\partial \varepsilon_t}{\partial u_{xx}} &= -\frac{\varepsilon_t^2 + \kappa_t^2 - 1}{2} + \frac{(\varepsilon_t - 1)^2 + \kappa_t^2}{2}\frac{\Pi}{\Lambda}\cos 2\phi_K, \\
\frac{\partial \mu_t}{\partial u_{xx}} &= -\frac{\mu_t^2 + \kappa_t^2 - 1}{2} + \frac{(\mu_t - 1)^2 + \kappa_t^2}{2}\frac{\Pi}{\Lambda}\cos 2\phi_K, \\
\frac{\partial \varepsilon_t}{\partial u_{xx}} &= -\frac{(\varepsilon_t + \mu_t)\kappa_t}{2} + \frac{(\varepsilon_t + \mu_t - 2)\kappa_t}{2}\frac{\Pi}{\Lambda}\cos 2\phi_K.
\end{aligned} \tag{S32}$$

The partial differentiate of the constitutive parameters with respect to $u_{yy}$ can be directly obtained by replacing $\Pi$ by $-\Pi$ [5].

For the sheared lattice, the lattice sums are

$$S_0 = i\Lambda \frac{1}{k_0^2 - K^2}, \quad S_{\pm 1} = \mp \Lambda \frac{K}{k_0(k_0^2 - K^2)} e^{\mp i\phi_K}, \quad S_{\pm 2} = -i\Lambda \frac{K^2}{k_0^2(k_0^2 - K^2)} e^{\mp 2i\phi_K} \mp \Xi u_{xy}, \tag{S33}$$

where $\Xi = 0.596\Lambda$ for the square lattice and $\Xi = -1.0\Lambda$ for the hexagonal lattice [5]. For the unit cell being sheared, substituting Eq. (S33) into the secular equation Eq. (S8), similarly we have

$$\frac{1}{k_0^4} \frac{\partial}{\partial u_{xy}} (K_+^2 K_-^2)\bigg|_{u_{xy} \to 0} = \frac{\partial}{\partial u_{xy}} (\varepsilon_z \mu_z - \kappa_z^2)(\varepsilon_t \mu_t - \kappa_t^2) = R \frac{U}{P^2}, \tag{S34}$$

with

$$U = -i\Lambda \Xi (A_1^2 + B_1^2 + 2\Lambda^2 A_1^2 B_1^2 - 4\Lambda^2 A_1 B_1 C_1^2 + 2C_1^2 + 2\Lambda^2 C_1^4). \tag{S35}$$

Then we can easily obtain that

$$\frac{\partial}{\partial u_{xy}} (\varepsilon_z \mu_z - \kappa_z^2) = 0, \quad \frac{\partial}{\partial u_{xy}} (\varepsilon_t \mu_t - \kappa_t^2) = \frac{U}{P^2} \tag{S36}$$

We can do the same procedures as we did for the tensors of diagonal term and finally have

$$\frac{\partial \varepsilon_z}{\partial u_{xy}} = \frac{\partial \mu_z}{\partial u_{xy}} = \frac{\partial \kappa_z}{\partial u_{xy}} = 0,$$

$$\frac{\partial \varepsilon_t}{\partial u_{xy}} = -\frac{(\varepsilon_t - 1)^2 + \kappa_t^2}{2} \frac{\Xi}{\Lambda} \sin 2\phi_K,$$

$$\frac{\partial \mu_t}{\partial u_{xy}} = -\frac{(\mu_t - 1)^2 + \kappa_t^2}{2} \frac{\Xi}{\Lambda} \sin 2\phi_K, \tag{S37}$$

$$\frac{\partial \kappa_t}{\partial u_{xy}} = -\frac{(\varepsilon_t + \mu_t - 2)\kappa_t}{2} \frac{\Xi}{\Lambda} \sin 2\phi_K,$$

The validity of the derived electo/magneto-strictive tensors in Eqs. (S32) and (S37) can be tested by comparing them with numerical simulation results, which are obtained using the eigen-fields and the band dispersions combined with finite-differences method, see detail in the following section.

**Section II: Numerical calculation of the effective constitutive parameters and their electro/magneto-strictive tensors**

**A. Numerical method based on bands and eigenfields**

The effective constitutive parameters of a photonic crystal can be obtained numerically by analyzing the eigen-fields and the band dispersions in the long wavelength limit. For photonic crystals with chiral inclusions, there are two kinds of eigen-fields corresponding to the lowest two bands. For each kind of eigen-fields, its spatial average $\bar{\mathbf{E}}_j = 1/\Omega \int_\Omega \mathbf{E}_j dxdy, \bar{\mathbf{H}}_j = 1/\Omega \int_\Omega \mathbf{H}_j dxdy$ should fulfill the Maxwell equations [7]

$$\nabla \times (\bar{\mathbf{E}}_j e^{ik_j x}) = i\omega \bar{\mathbf{B}}_j e^{ik_j x} = i\omega(\vec{\mu}\bar{\mathbf{H}}_j - i\vec{\kappa}\bar{\mathbf{E}}_j)e^{ik_j x},$$
$$\nabla \times (\bar{\mathbf{H}}_j e^{ik_j x}) = -i\omega \bar{\mathbf{D}}_j e^{ik_j x} = -i\omega(\vec{\varepsilon}\bar{\mathbf{E}}_j + i\vec{\kappa}\bar{\mathbf{H}}_j)e^{ik_j x}, \tag{S38}$$

where $k_j, j=1,2$ is the magnitude of Bloch vector for each corresponding eigen-field which can be complex numbers, $\hat{x}$ denotes the direction of the Bloch vector, $\vec{\varepsilon}, \vec{\mu}$ and $\vec{\kappa}$ are the effective constitutive parameters defined in Eq. (1) in main text. Then Eq. (S38) is reduced to

$$ik_j \begin{pmatrix} 0 \\ -E_{zj} \\ E_{yj} \end{pmatrix} = i\omega(\begin{pmatrix} 0 \\ \mu_t H_{yj} \\ \mu_z H_{zj} \end{pmatrix} - i \begin{pmatrix} 0 \\ \kappa_t E_{yj} \\ \kappa_z E_{zj} \end{pmatrix}), \quad ik_j \begin{pmatrix} 0 \\ -H_{zj} \\ H_{yj} \end{pmatrix} = -i\omega(\begin{pmatrix} 0 \\ \varepsilon_t E_{yj} \\ \varepsilon_z E_{zj} \end{pmatrix} + i \begin{pmatrix} 0 \\ \kappa_t H_{yj} \\ \kappa_z H_{zj} \end{pmatrix}), \tag{S39}$$

where

$$\bar{\mathbf{E}}_j = \frac{1}{\Omega}\int_\Omega \mathbf{E}_j dxdy = \begin{pmatrix} 0 \\ E_{yj} \\ E_{zj} \end{pmatrix}, \qquad \bar{\mathbf{H}}_j = \frac{1}{\Omega}\int_\Omega \mathbf{H}_j dxdy = \begin{pmatrix} 0 \\ H_{yj} \\ H_{zj} \end{pmatrix}, \tag{S40}$$

Solving Eq. (S39), the constitutive parameters can be obtained as

$$\varepsilon_z = \frac{n_1 H_{y1} H_{z2} - n_2 H_{y2} H_{z1}}{E_{z2} H_{z1} - E_{z1} H_{z2}}, \quad \varepsilon_t = -\frac{n_1 H_{y2} H_{z1} - n_2 H_{y1} H_{z2}}{E_{y2} H_{y1} - E_{y1} H_{y2}}, \tag{S41a}$$

$$\kappa_z = i\frac{n_1 H_{y1} E_{z2} - n_2 H_{y2} E_{z1}}{E_{z2} H_{z1} - E_{z1} H_{z2}}, \quad \kappa_t = -i\frac{n_1 E_{y2} H_{z1} - n_2 E_{y1} H_{z2}}{E_{y2} H_{y1} - E_{y1} H_{y2}}, \tag{S41b}$$

$$\mu_z = \frac{n_1 E_{y1} E_{z2} - n_2 E_{y2} E_{z1}}{E_{z2} H_{z1} - E_{z1} H_{z2}}, \quad \mu_t = -\frac{n_1 E_{y2} E_{z1} - n_2 E_{y1} E_{z2}}{E_{y2} H_{y1} - E_{y1} H_{y2}}, \tag{S41c}$$

$$\kappa_z = -i\frac{n_1 E_{y1} H_{z2} - n_2 E_{y2} H_{z1}}{E_{z2} H_{z1} - E_{z1} H_{z2}}, \quad \kappa_t = i\frac{n_1 H_{y2} E_{z1} - n_2 H_{y1} E_{z2}}{E_{y2} H_{y1} - E_{y1} H_{y2}}, \tag{S41d}$$

where $n_1 = k_1/\omega, n_2 = k_2/\omega$ in the limit $k_{1,2} \to 0, \omega \to 0$ are the effective refractive indices of the metamaterial. The two equations Eq. (S41b) and Eq. (S41d) about the chirality tensors always give the same result due to the relation between the two eigen-fields.

We can repeat the calculation after deforming the unit cell, and the electrostrictive and magnetostrictive tensors can be obtained using the finite differences. For example, for the square lattice, we obtain the effective permittivity of the out-plane component before and after the unit cell being stretched using the Eq. (S41) as $\varepsilon_z$ and $\varepsilon_z'$, then the electrostrictive term is given by

$$\frac{\partial \varepsilon_z}{\partial u_{xx}} = \frac{\varepsilon_z' - \varepsilon_z}{(2\Delta a / a)}$$

where $a$ is the lattice constant, and $\Delta a$ is an infinitesimal stretching displacement along the $x$ direction, see Fig.S3.

## B. Calculating complex K Bloch bands for the artificial chiral inclusions

To obtain the eigen-fields and the corresponding Bloch K vectors, we need to calculate the complex K Bloch bands, which can be calculated using the Weak-Form-PDE module in COMSOL [12-14].

For bi-anisotropic medium possesses the following constitutive relations

$$\mathbf{D} = \ddot{\varepsilon}\varepsilon_0 \mathbf{E} + i\vec{\kappa}/c\mathbf{H}, \quad \mathbf{B} = \ddot{\mu}\mu_0 \mathbf{H} - i\vec{\kappa}/c\mathbf{E}, \tag{S42}$$

and if the constitutive parameters have diagonal matrix forms, the wave equations inside the medium are

$$\nabla \times \left[ (\ddot{\mu}^{-1}\ddot{\kappa} - \ddot{\kappa}^{-1}\ddot{\varepsilon})^{-1}(i\ddot{\mu}^{-1}\nabla \times \mathbf{E} + c\ddot{\kappa}^{-1}\nabla \times \mathbf{H}) \right] = -\frac{\omega^2}{c^2}(\ddot{\mu}\mathbf{H} - i\ddot{\kappa}\mathbf{E}),$$

$$\nabla \times \left[ (\ddot{\kappa}^{-1}\ddot{\mu} - \ddot{\varepsilon}^{-1}\ddot{\kappa})^{-1}(\ddot{\kappa}^{-1}\nabla \times \mathbf{E} - i\ddot{\varepsilon}^{-1}\nabla \times \mathbf{H}) \right] = \frac{\omega^2}{c^2}(\ddot{\varepsilon}\mathbf{E} + i\ddot{\kappa}\mathbf{H}). \tag{S43}$$

For the 2D system, because the transverse fields can be obtained from the z component fields according to the Maxwell equations, we only consider the z component fields. And using the Bloch theorem

$$E_z = u(\mathbf{r})\exp[-i(\omega t - \mathbf{k}\cdot\mathbf{r})], \quad H_z = v(\mathbf{r})\exp[-i(\omega t - \mathbf{k}\cdot\mathbf{r})], \tag{S44}$$

then the wave equations for z component fields are

$$(ik_x, ik_y) \times \left[ f_1(u_y + ik_y u, -u_x - ik_x u) - if_2(v_y + ik_y v, -v_x - ik_x v) \right]$$
$$+ \nabla \times \left[ f_1(u_y + ik_y u, -u_x - ik_x u) - if_2(v_y + ik_y v, -v_x - ik_x v) \right] - i\frac{\omega^2}{c^2}(\mu_z v - i\kappa_z u) = 0, \tag{S45a}$$

$$(ik_x, ik_y) \times \left[ if_3(u_y + ik_y u, -u_x - ik_x u) + f_1(v_y + ik_y v, -v_x - ik_x v) \right]$$
$$+ \nabla \times \left[ if_3(u_y + ik_y u, -u_x - ik_x u) + f_1(v_y + ik_y v, -v_x - ik_x v) \right] + i\frac{\omega^2}{c^2}(\varepsilon_z u + i\kappa_z v) = 0, \tag{S45b}$$

where

$$f_1 = \frac{\kappa_t}{\kappa_t^2 - \varepsilon_t\mu_t}, \quad f_2 = \frac{\mu_t}{\kappa_t^2 - \varepsilon_t\mu_t}, \quad f_3 = \frac{\varepsilon_t}{\kappa_t^2 - \varepsilon_t\mu_t}$$

and $u_x = \partial_x u$, $u_y = \partial_y u$, $v_x = \partial_x v$, $v_y = \partial_y v$. Multiply the test functions $\tilde{u}, \tilde{v}$, respectively, and integrate within a unit cell, we then obtain the weak forms as

$$Wk(u) = \left( f_1 k^2 u + if_1 k_x u_x + if_1 k_y u_y - if_2 k^2 v + f_2 k_x v_x + f_2 k_y v_y \right)\tilde{u}$$
$$- \left( if_1 k_x \tilde{u}_x u + if_1 k_y \tilde{u}_y u - f_1 \tilde{u}_x u_x - f_1 \tilde{u}_y u_y + f_2 k_x \tilde{u}_x v + f_2 k_y \tilde{u}_y v + if_2 \tilde{u}_x v_x + if_2 \tilde{u}_y v_y \right) \tag{S46a}$$
$$- i\frac{\omega^2}{c^2}(\mu v - i\kappa u)\tilde{u} = 0,$$

$$Wk(v) = \left(if_3k^2u - f_3k_xu_x - f_3k_yu_y + f_1k^2v + if_1k_xv_x + if_1k_yv_y\right)\tilde{v}$$
$$-\left(-f_3k_x\tilde{v}_xu - f_3k_y\tilde{v}_yu - if_3\tilde{v}_xu_x - if_3\tilde{v}_yu_y + if_1k_x\tilde{v}_xv + if_1k_y\tilde{v}_yv - f_1\tilde{v}_xv_x - f_1\tilde{v}_yv_y\right) \quad \text{(S46b)}$$
$$+i\frac{\omega^2}{c^2}(\varepsilon u + i\kappa v)\tilde{v} = 0$$

In the simulations, the fields should be approximated with Lagrange interpolation elements. The transverse component fields can be obtained according to the Maxwell equations. Then knowing the complex Bloch K bands and the EM fields, we can numerically calculate the effective constitutive parameters and the electro/magneto-strictive tensors following the method proposed in Sec. II. A.

## C. The electro/magneto-strictive tensors for the metamaterials composed of artificial chiral cylinders

For bi-anisotropic metamaterials composed of type I or type II cylinders (see text), we numerically calculate the electro/magnetostrictive tensors according to Eq. (S41) and the finite difference method. The numerical results compared with the formula results are shown in Fig. S4 and S5. We can see that within the numerical error, the numerical results accord with the formula results very well, indicating that Eq. (7) in the main text can be used for calculating the electro/magneto-strictive tensors for these bi-anisotropic metamaterials correctly.

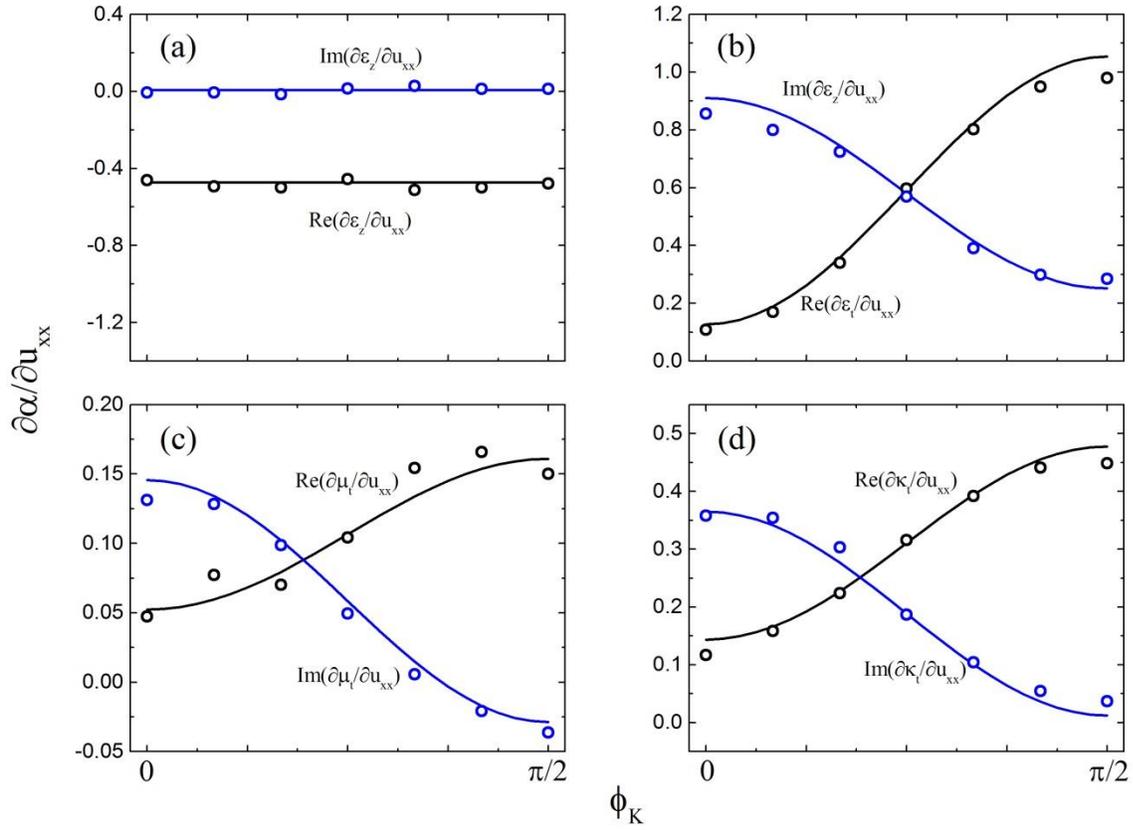

**Fig. S4**. For metamaterials composed by type I chiral cylinders, the comparison between tensor components obtained from formulas (lines) and those from numerical calculations (circles). Results as functions of the direction of Bloch vector $\phi_K$ obtained from the formulas and numerical calculations are shown by lines and circles, respectively.

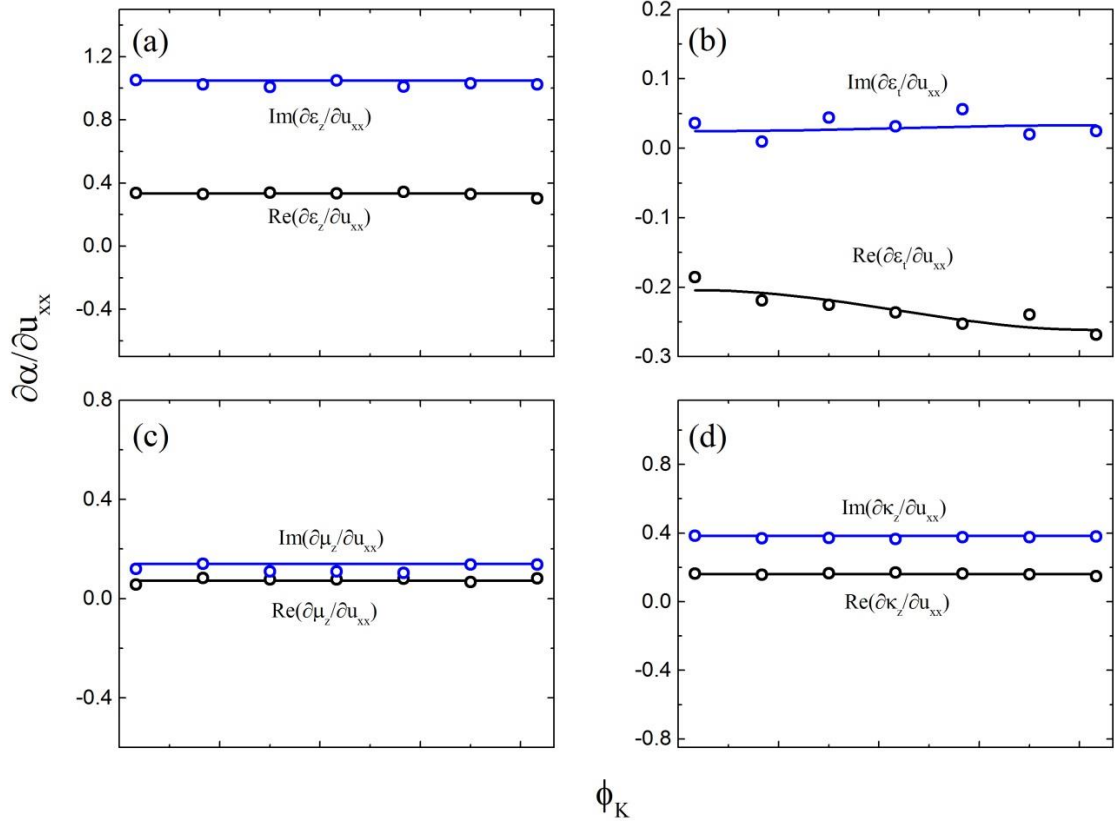

**Fig. S5**. For metamaterials composed by type II chiral cylinders, the comparison between tensor components obtained from formulas (lines) and those from numerical calculations (circles). Results as functions of the direction of Bloch vector $\phi_K$ obtained from the formulas and numerical calculations are shown by lines and circles, respectively.

## Section III: Derivation of the extended stress tensor for bi-anisotropic medium

Electromagnetic stress tensors can be obtained using the virtual work principle under the quasi static limit [9, 15]. Conventional stress tensors are formulated for achiral medium. Here we derive an extended form of the stress tensor that works for the bi-anisotropic medium whose constitutive parameters have the following forms

$$\ddot{\varepsilon}_e = \varepsilon_t \hat{x}\hat{x} + \varepsilon_t \hat{y}\hat{y} + \varepsilon_z \hat{z}\hat{z}, \quad \ddot{\mu}_e = \mu_t \hat{x}\hat{x} + \mu_t \hat{y}\hat{y} + \mu_z \hat{z}\hat{z}, \quad \ddot{\kappa}_e = \kappa_t \hat{x}\hat{x} + \kappa_t \hat{y}\hat{y} + \kappa_z \hat{z}\hat{z}. \tag{S47}$$

Consider a small square area inside the bi-anisotropic medium with volume $ds = a \times b$, as shown in Fig. S1. As the electromagnetic (EM) fields inside this area are almost constant in the long wavelength limit, the time-averaged total EM energy for this area is given by $W = -\frac{1}{4}\text{Re}(\mathbf{E}\cdot\mathbf{D}^* + \mathbf{H}\cdot\mathbf{B}^*)ab$. If we subject one of the boundaries to a virtual translation over an infinitesimal distance $\boldsymbol{\xi}$, then the variance of the total electric energy $\delta W$ should be just equal to the work done by the boundary force $\sum T_{ik}\xi_i n_k b$, where $T_{ik}$ is the surface stress tensor and $\mathbf{n}$ is the unit normal vector of the boundary. Hence we have

$$\sum T_{ik}\xi_i n_k b = \delta W = \delta W_s + \delta W_f + \delta W_p, \tag{S48}$$

where the variation of total EM energy $\delta W$ consists of three parts which are due to the variations of total volume of the area, EM fields and constitutive parameters, respectively, and they are given by

$$\delta W_s = -\frac{1}{4}\text{Re}(\mathbf{E}\cdot\mathbf{D}^* + \mathbf{H}\cdot\mathbf{B}^*)b\mathbf{n}\cdot\boldsymbol{\xi} = -\frac{1}{4}\text{Re}(\mathbf{E}\cdot\mathbf{D}^* + \mathbf{H}\cdot\mathbf{B}^*)b\sum\delta_{ik}\xi_i n_k, \tag{S49a}$$

$$\delta W_f = \frac{\partial}{\partial \mathbf{E}}[-\frac{1}{4}\text{Re}(\mathbf{E}\cdot\mathbf{D}^* + \mathbf{H}\cdot\mathbf{B}^*)]\cdot\delta\mathbf{E}ab + \frac{\partial}{\partial \mathbf{H}}[-\frac{1}{4}\text{Re}(\mathbf{E}\cdot\mathbf{D}^* + \mathbf{H}\cdot\mathbf{B}^*)]\cdot\delta\mathbf{H}ab, \tag{S49b}$$

$$\delta W_p = \sum \frac{\partial}{\partial \eta}[-\frac{1}{4}\text{Re}(\mathbf{E}\cdot\mathbf{D}^* + \mathbf{H}\cdot\mathbf{B}^*)]\cdot\delta\eta ab, \tag{S49c}$$

where $\delta_{ik}$ is Kronecher delta function, $\eta = \varepsilon_t, \varepsilon_z, \mu_t, \mu_z, \kappa_t, \kappa_z$ are constitutive parameters. Note that the potential of each point on the boundary remains invariant during the deformation [9], namely $\mathbf{E}'\cdot\mathbf{n}a + \mathbf{E}'\cdot\boldsymbol{\xi} = \mathbf{E}\cdot\mathbf{n}a$ and $\mathbf{E}'\times\mathbf{n}b = \mathbf{E}\times\mathbf{n}b$, then we have

$$\delta\mathbf{E} = \mathbf{E}' - \mathbf{E} = -\mathbf{n}\frac{\mathbf{E}\cdot\boldsymbol{\xi}}{a}, \qquad \delta\mathbf{H} = \mathbf{H}' - \mathbf{H} = -\mathbf{n}\frac{\mathbf{H}\cdot\boldsymbol{\xi}}{a}, \tag{S50}$$

Substituting the relations into Eq. (49b), we have

$$\delta W_f = \frac{1}{2}\text{Re}(\mathbf{n}\cdot\mathbf{D}^*)(\mathbf{E}\cdot\boldsymbol{\xi})b + \frac{1}{2}\text{Re}(\mathbf{n}\cdot\mathbf{B}^*)(\mathbf{H}\cdot\boldsymbol{\xi})b = \frac{1}{2}\sum\text{Re}(E_i D_k^* + H_i B_k^*)\xi_i n_k. \tag{S51}$$

The constitutive parameters are related to the strain tensors,

$$\delta\varepsilon_j = \frac{\partial\varepsilon_j}{\partial u_{ik}} u_{ik}, \qquad \delta\mu_j = \frac{\partial\mu_j}{\partial u_{ik}} u_{ik}, \qquad \delta\kappa_j = \frac{\partial\kappa_j}{\partial u_{ik}} u_{ik}.$$

with the strain tensors given by [9]

$$u_{ik} = \frac{1}{2}\left(\frac{\partial u_i}{\partial x_k} + \frac{\partial u_k}{\partial x_j}\right) = \frac{1}{2a}(\xi_i n_k + \xi_k n_i).$$

Then Eq.(S49c) is reduced to

$$\delta W_p = -\frac{1}{4}\mathrm{Re}\left[\begin{array}{l} \varepsilon_0(\frac{\partial\varepsilon_t}{\partial u_{ik}}|\mathbf{E}_t|^2 + \frac{\partial\varepsilon_z}{\partial u_{ik}}|E_z|^2) + \mu_0(\frac{\partial\mu_t}{\partial u_{ik}}|\mathbf{H}_t|^2 + \frac{\partial\mu_z}{\partial u_{ik}}|H_z|^2) \\ +\frac{2}{c}\mathrm{Im}(E_x H_x^* + E_y H_y^*)\frac{\partial\kappa_t}{\partial u_{ik}} + \frac{2}{c}\mathrm{Im}(E_z H_z^*)\frac{\partial\kappa_z}{\partial u_{ik}} \end{array}\right]. \tag{S52}$$

Where $\mathbf{E}_t = (E_x, E_y), \mathbf{H}_t = (H_x, H_y)$ are EM fields in the *xy* plane. Combining Eqs. (S48), (S49a), (S51) and (S52), then we can generalize the extended Helmholtz stress tensor for the bi-anisotropic medium as

$$T_{ik} = \frac{1}{2}\mathrm{Re}\left\{\begin{array}{l} E_i D_k^* + H_i B_k^* - \frac{1}{2}(\mathbf{E}\cdot\mathbf{D}^* + \mathbf{H}\cdot\mathbf{B}^*)\delta_{ik} - \frac{1}{2}\varepsilon_0(\frac{\partial\varepsilon_t}{\partial u_{ik}}|\mathbf{E}_t|^2 + \frac{\partial\varepsilon_z}{\partial u_{ik}}|E_z|^2) \\ -\frac{1}{2}\mu_0(\frac{\partial\mu_t}{\partial u_{ik}}|\mathbf{H}_t|^2 + \frac{\partial\mu_z}{\partial u_{ik}}|H_z|^2) + \frac{1}{c}\mathrm{Im}(E_x H_x^* + E_y H_y^*)\frac{\partial\kappa_t}{\partial u_{ik}} + \frac{1}{c}\mathrm{Im}(E_z H_z^*)\frac{\partial\kappa_z}{\partial u_{ik}} \end{array}\right\}. \tag{S53}$$

It is clearly that Eq. (S53) reduces to the traditional Helmhlotz stress tensor when $\kappa_t = \kappa_z = 0$ and Maxwell stress tensor when the medium is vacuum.

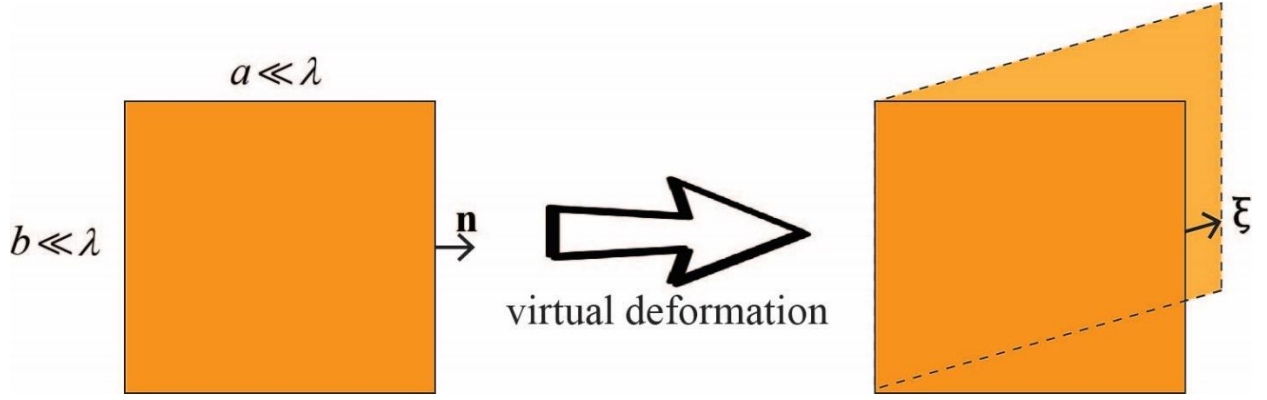

**Fig. S6.. Derivation of Helmholtz stress tensor by virtual work principle.** For a very small area *dS*, the total electric energy of this area is $-1/4\,\text{Re}(\mathbf{E}\cdot\mathbf{D}^* + \mathbf{H}\cdot\mathbf{B}^*)dS$. If we subject one of the boundaries to a virtual translation over an infinitesimal distance $\boldsymbol{\xi}$, then the variance of the total electric energy should be just equal to the work done by the electric component of the boundary force $T_{ik}\xi_i n_k b$, where $T_{ik}$ is the surface stress tensor and $\mathbf{n}$ is the normal vector of the boundary.

### Section IV. Numerically obtaining the Mie coefficients of the artificial cylinders

The Mie coefficients of the artificial chiral cylinder can be determined according to the scattering fields $\mathbf{E}_s, \mathbf{H}_s$ of the cylinder in the far field region. Since the scattering fields can be written as series of vector cylindrical wave functions (VCWFs) $\mathbf{M}_n^{(J)}, \mathbf{N}_n^{(J)}$, see Eq. (S1), and the VCWFs are orthogonal to each other, the scattering coefficients can be obtained by

$$a_n = \frac{1}{2\pi H_{n-1}^{(1)}(k_0 r) H_{n+1}^{(1)}(k_0 r)} \int_0^{2\pi} \mathbf{E}_s \cdot [\mathbf{M}_n^{(3)}(k_0, \mathbf{r})]^* d\phi,$$

$$b_n = -\frac{1}{2\pi [H_n^{(1)}(k_0 r)]^2} \int_0^{2\pi} \mathbf{E}_s \cdot [\mathbf{N}_n^{(3)}(k_0, \mathbf{r})]^* d\phi. \tag{S54}$$

where

$$[\mathbf{M}_n^{(3)}(k, \mathbf{r})]^* = [\frac{in}{kr} H_n^{(1)}(kr)\mathbf{e}_r - H_n^{(1)}{}'(kr)\mathbf{e}_\phi]e^{-in\phi},$$

$$[\mathbf{N}_n^{(3)}(k, \mathbf{r})]^* = H_n^{(1)}(kr)e^{-in\phi}\mathbf{e}_z.$$

The scattering coefficients can also be obtained according to the magnetic scattering field as

$$a_n = \frac{1}{i} \frac{1}{2\pi [H_n^{(1)}(k_0 r)]^2} \int_0^{2\pi} \mathbf{H}_s \cdot [\mathbf{N}_n^{(3)}(k_0, \mathbf{r})]^* d\phi,$$

$$b_n = \frac{i}{2\pi H_{n-1}^{(1)}(k_0 r) H_{n+1}^{(1)}(k_0 r)} \int_0^{2\pi} \mathbf{H}_s \cdot [\mathbf{M}_n^{(3)}(k_0, \mathbf{r})]^* d\phi. \tag{S55}$$

Eq. (S54) and Eq. (S55) actually give almost the same results in the numerical calculations. And then we can further obtain the Mie coefficients according to the relations

$$a_n = A_n p_n + C_n q_n, \qquad b_n = C_n p_n + B_n q_n.$$

For Hz polarized plane wave propagating along the $x$ direction, we have $p_n = -i^n, q_n = 0$, and for Ez polarized plane wave propagating along the $x$ direction, we have $p_n = 0, q_n = i^n$. As a result, we can obtain all the Mie coefficients after solving the scattering fields in the far field region when the artificial cylinder illuminated by a Hz and a Ez polarized plane wave and using Eq. (S54) and Eq. (S55). In the simulations, we use the commercial software COMSOL [8] to solve the scattering fields of a single artificial cylinder and then using the Eq. (S54) and Eq. (S55) to calculate the Mie coefficients. And we found that the Mie coefficients satisfy $A_n \approx A_{-n}, B_n \approx B_{-n}, C_n \approx C_{-n}$, thus the effective constitutive parameters of the matematerials composed by the artificial cylinders can be well described by Eq. (1) in the main text.

**Section V. Spatially averaged fields and energy density inside the metamaterials**

For type I bi-anisotropic metamaterial, we show the consistencies between the spatially averaged lattice fields and the fields inside the effective medium for $x$ and $z$ components in Fig. S7.

In Fig. S8 we show that for both the type I and type II bi-anisotropic metamaterials, the spatially averaged energy densities inside the metamaterial using full wave simulations agree with the energy densities inside the corresponding effective mediums very well. These consistencies provide the necessary condition for applying the extended Helmholtz stress tensor to calculate the optical force density. We note that the agreements are not good near the boundaries. It is expected since any type of EMT description will fail close to the boundaries.

We also note that the spatially averaged energy density is not exactly identical to the corresponding energy density inside the effective medium, this is because the lattice constant in the real structure is not so small compared with the wavelength. In the following section, we will analytically show that these two energy densities are equal in the true long-wavelength limit.

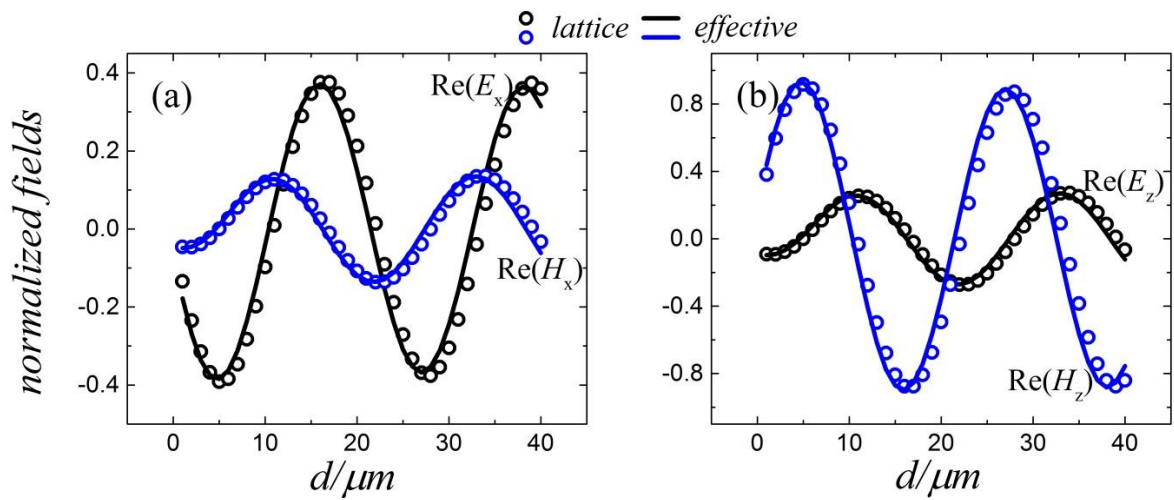

**Figure S7.** For x and z components, comparison between the spatially averaged EM fields inside bi-anisotropic metamaterials using full wave simulations and the EM fields in the corresponding effective medium. (a) The parameters used are the same with those in Fig. 2.

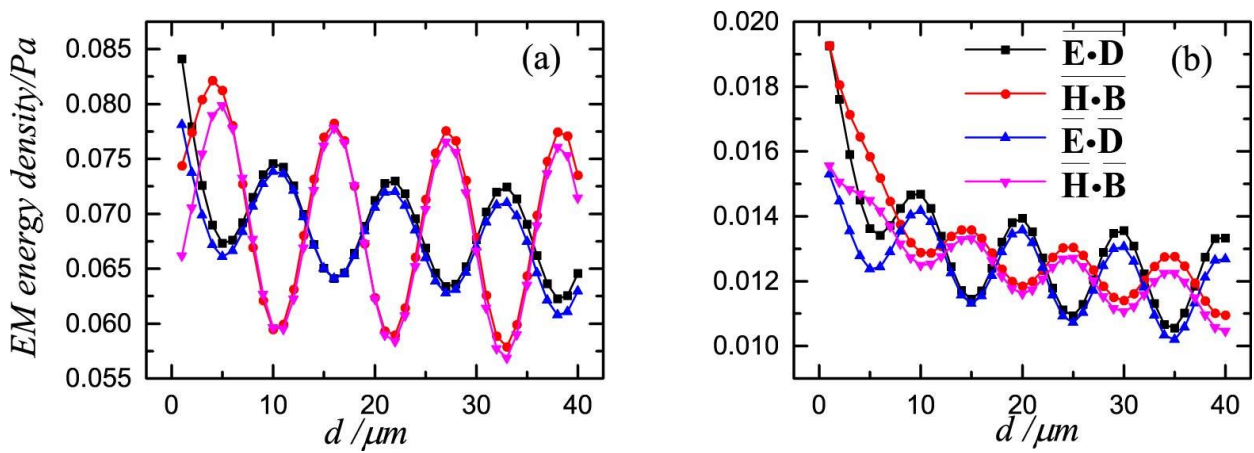

**Figure S8.** Comparison between the spatially averaged EM energy density inside bi-anisotropic metamaterials using full wave simulations and the EM energy density in the corresponding effective medium. (a) The parameters used are the same with those in Fig. 2. (b) The parameters used are the same with those in Fig. 3(a)-(b).

## Section VI. The equality relationship between the microscopic and macroscopic energy densities

In this section, we will show that from the spatial average relation between the microscopic and macroscopic fields, the equality relationship between the microscopic and macroscopic energy densities can be obtained.

In the long wavelength limit, the effective fields (macroscopic fields) inside the effective medium are defined as the spatial average of the fields (microscopic fields) inside the metamaterial [5], namely

$$\mathbf{E}_e = \frac{1}{\Omega}\int_\Omega \mathbf{E} d\Omega, \qquad \mathbf{D}_e = \frac{1}{\Omega}\int_\Omega \mathbf{D} d\Omega, \tag{S56}$$

where $\Omega$ is the volume of the unit cell, $\mathbf{E}, \mathbf{D}$ and $\mathbf{E}_e, \mathbf{D}_e$ are the fields inside the metamaterial and effective medium, respectively. Starting from these relations, we will show that the energy density in the effective medium is also equal to the spatial average of the energy density inside the metamaterial, namely

$$-\frac{1}{4}\mathbf{E}_e \cdot \mathbf{D}_e^* = -\frac{1}{4\Omega}\int_\Omega \mathbf{E}\cdot\mathbf{D}^* d\Omega. \tag{S57}$$

For the field in the metamaterial, according to the Maxwell equations and the long-wavelength limit where $\omega \to 0, k \to 0$, we have

$$\nabla \times \mathbf{E} = i\omega\mathbf{B} \approx 0. \tag{S58}$$

That is the electric field is a curl-free vector, therefore it can be written as gradient of a scalar,

$$\mathbf{E} = \nabla\phi. \tag{S59}$$

Substituting the relations into Eq. (S56), we have

$$\mathbf{E}_e = \frac{1}{\Omega}\int_\Omega \nabla\phi = \frac{1}{\Omega}\oint_\Gamma \phi\hat{n}d\Gamma, \tag{S60}$$

where $\Gamma$ is the boundary of the unit cell and $\hat{n}$ is the unit normal vector of the boundary. And for the spatial average of electric energy in a unit cell, we have

$$-\frac{1}{4\Omega}\int_\Omega \mathbf{E}\cdot\mathbf{D}^* d\Omega = -\frac{1}{4\Omega}\int_\Omega \nabla\phi\cdot\mathbf{D}^* d\Omega = -\frac{1}{4\Omega}\int_\Omega [\nabla\cdot(\phi\mathbf{D}^*) - \phi\nabla\cdot\mathbf{D}^*]d\Omega$$
$$= -\frac{1}{4\Omega}\int_\Omega [\nabla\cdot(\phi\mathbf{D}^*)]d\Omega = -\frac{1}{4\Omega}\int_\Omega \phi\mathbf{D}^*\cdot\hat{n}d\Gamma, \tag{S61}$$

where the Maxwell equation $\nabla\cdot\mathbf{D}=0$ is used in the derivation.

In the following, we will take the cubic lattice as an example, other lattices can be followed in the same way. For the cubic lattice with lattice constant $a$, according to Eq. (S60), the electric field of each direction is given by

$$E_{ex} = \frac{1}{\Omega}\oint_\Gamma \phi\hat{e}_x\cdot\hat{n}d\Gamma = \frac{1}{a^3}\int_0^a\int_0^a [\phi(a,y,z)-\phi(0,y,z)]dydz,$$
$$E_{ey} = \frac{1}{\Omega}\oint_\Gamma \phi\hat{e}_y\cdot\hat{n}d\Gamma = \frac{1}{a^3}\int_0^a\int_0^a [\phi(x,a,z)-\phi(x,0,z)]dxdz, \tag{S62}$$
$$E_{ez} = \frac{1}{\Omega}\oint_\Gamma \phi\hat{e}_z\cdot\hat{n}d\Gamma = \frac{1}{a^3}\int_0^a\int_0^a [\phi(x,y,a)-\phi(x,y,0)]dxdz.$$

And note that

$$\frac{\partial}{\partial y}[\phi(a,y,z)-\phi(0,y,z)] = E_y(a,y,z) - E_y(0,y,z) = E_y(0,y,z)(e^{ik_x a}-1) = 0,$$
$$\frac{\partial}{\partial z}[\phi(a,y,z)-\phi(0,y,z)] = E_z(a,y,z) - E_z(0,y,z) = E_z(0,y,z)(e^{ik_x a}-1) = 0,$$

where the long-wavelength limit that $k_x \to 0, k_y \to 0, k_z \to 0$ are used in the derivations. It means that the functions in the brackets of Eq. (S62) are independent with the coordinates and can be picked out. Thus the effective electric field can be rewritten as

$$E_{ex} = \frac{1}{a^3}\int_0^a\int_0^a [\phi(a,y,z) - \phi(0,y,z)]dydz = \frac{1}{a}[\phi(a,y,z) - \phi(0,y,z)] = \frac{1}{a}\int_0^a E_x dx,$$

$$E_{ey} = \frac{1}{a^3}\int_0^a\int_0^a [\phi(x,a,z) - \phi(x,0,z)]dxdz = \frac{1}{a}[\phi(x,a,z) - \phi(x,0,z)] = \frac{1}{a}\int_0^a E_y dy, \quad (S63)$$

$$E_{ez} = \frac{1}{a^3}\int_0^a\int_0^a [\phi(x,y,a) - \phi(x,y,0)]dxdz = \frac{1}{a}[\phi(x,y,a) - \phi(x,y,0)] = \frac{1}{a}\int_0^a E_z dz.$$

And for the electric displacement, take the $D_{ex}$ as an example, we use the integration by part that

$$\begin{aligned}
D_{ex} &= \frac{1}{a^3}\iiint_\Omega D_x(x,y,z)dxdydz = \frac{1}{a^3}x\iint D_x(x,y,z)dydz\Big|_0^a - \frac{1}{a^3}\int_0^a [x\frac{\partial}{\partial x}\iint D_x(x,y,z)dydz]dx \\
&= \frac{1}{a^2}\iint D_x(a,y,z)dydz - \frac{1}{a^3}\int_0^a [x\iint \frac{\partial}{\partial x}D_x(x,y,z)dydz]dx \\
&= \frac{1}{a^2}\iint D_x(a,y,z)dydz + \frac{1}{a^3}\int_0^a \{x\iint [\frac{\partial}{\partial y}D_y(x,y,z) + \frac{\partial}{\partial z}D_z(x,y,z)]dydz\}dx \quad (S64) \\
&= \frac{1}{a^2}\iint D_x(a,y,z)dydz + \frac{1}{a^3}\int_0^a \{x\int_0^a [D_y(x,a,z) - D_y(x,0,z)]dz\}dx \\
&\quad + \frac{1}{a^3}\int_0^a \{x\int_0^a [D_z(x,y,a) - D_z(x,y,0)]dy\}dx = \frac{1}{a^2}\iint D_x(a,y,z)dydz.
\end{aligned}$$

Here we used that $\nabla\cdot\mathbf{D} = \partial D_x/\partial x + \partial D_y/\partial y + \partial D_z/\partial z = 0$ and $\mathbf{D}(x,y,a) = \mathbf{D}(x,y,0)e^{ik_z a} = \mathbf{D}(x,y,0)$ in the derivations. The same procedure can be done for other components. Then according to Eq. (S63) and Eq. (S64), we can summarized that

$$E_{ei} = \frac{1}{a}\int_0^a \mathbf{E}\cdot d\mathbf{x}_i, \qquad D_{ei} = \frac{1}{a^2}\iint \mathbf{D}\cdot d\mathbf{S}_i, \quad (S65)$$

which is just the homogenization theory proposed by J. B. Pendry [16].

Substituting the relations Eq. (S65) into Eq. (S61), we can further obtain the relation between the electric energy densities,

$$-\frac{1}{4\Omega}\int_\Omega \mathbf{E}\cdot\mathbf{D}^* d\Omega = -\frac{1}{4\Omega}\int_\Omega \phi \mathbf{D}^*\cdot\hat{n}d\Gamma$$

$$= -\frac{1}{4a^3}\left\{\begin{array}{l}\int_0^a\int_0^a[\phi(a,y,z)D_x^*(a,y,z)-\phi(0,y,z)D_x^*(0,y,z)]dydz\\+\int_0^a\int_0^a[\phi(x,a,z)D_y^*(x,a,z)-\phi(x,0,z)D_y^*(x,0,z)]dxdz\\+\int_0^a\int_0^a[\phi(x,y,a)D_z^*(x,y,a)-\phi(x,y,0)D_z^*(x,y,0)]dxdy\end{array}\right\}$$

$$= -\frac{1}{4a^3}\left\{\begin{array}{l}[\phi(a,y,z)-\phi(0,y,z)]\int_0^a\int_0^a D_x^*(a,y,z)dydz\\+[\phi(x,a,z)-\phi(x,0,z)]\int_0^a\int_0^a D_y^*(x,a,z)dxdz\\+[\phi(x,y,a)-\phi(x,y,0)]\int_0^a\int_0^a D_z^*(x,y,a)dxdy\end{array}\right\} \quad (S66)$$

$$= -\frac{1}{4}(E_{ex}D_{ex}^* + E_{ey}D_{ey}^* + E_{ez}D_{ez}^*) = -\frac{1}{4}\mathbf{E}_e\cdot\mathbf{D}_e^*,$$

which is just what we desired. Similarly, we can also obtain

$$-\frac{1}{4\Omega}\int_\Omega \mathbf{H}\cdot\mathbf{B}^* d\Omega = -\frac{1}{4}\mathbf{H}_e\cdot\mathbf{B}_e^*,$$

for the magnetic component.

**References.**

1. C. F. Bohren and D. R. Huffman, *Absorption and Scattering of Light by Small Particles* (Wiley and Sons, 1983).

2. M. Abramowitz and I. A. Stegun, *Handbook of Mathematical Functions with Formulas, Graphs, and Mathematical Tables* (Wiley, New York, 1972).

3. Y. Wu and Z. Q. Zhang, *Phys. Rev. B* **79**, 195111, 2009.

4. S. K. Chin, N. A. Nicorovici, and R. C. McPhedran, *Phys. Rev. E* **49**, 4590 (1994).

5. W. Sun, S. B. Wang, J. Ng, L. Zhou and C. T. Chan, *Phys. Rev. B* **91**, 235439 (2015).

6. A. Lakhtakia, V. K. Varadan and V. V. Varadan, *J. Mater. Res.* **8**, 917 (1992).

7. M. G. Silveirinha, *Phys. Rev. B* **75**, 115104 (2007).

8. www.comsol.com.

9. L. D. Landau, E. M. Lifshitz, and L. P. Pitaevskii, *Electrodynamics of Continuous Media*, 2nd ed. (Butterworth-Heinemann, New York, 1984).

10. P. Penfield and H. A. Haus, *Electrodynamics of Moving Media* (MIT Press, Cambridge, 1967).

11. L. D. Landau and E. M. Lifshitz, *Theory of Elasticity*, 3rd ed. (Butterworth-Heinemann, New York, 1986).

12. M. Davanc, Y. Urzhumov, G. Shvets, *Opt. Express* **15**, 9681 (2007).

13. M. Davanc, Y. Urzhumov, G. Shvets, *Opt. Express* **19**, 19027 (2011).

14. J.-M. Jin, *The Finite Element Method in Electromagnetics* (John Wiley &Sons, Hoboken, 2014).

15. H. Helmholtz, *Ann. Phys.* **249**, 385 (1881).

16. D. R. Smith and J. B. Pendry, *J. Opt. Soc. Am. B.* **23**, 391 (2006).